\documentclass[aps,prd,superscriptaddress,groupedaddress,secnumarabic,floatfix,tightenlines,nobibnotes, 12pt,]{revtex4} 
\usepackage{blindtext}
\usepackage{threeparttable}
\usepackage{tikz}
\usepackage{spverbatim}
 
\usepackage[section]{placeins}
 \usepackage{subcaption}
\usepackage{listings}

\usepackage{rotating}
\newcommand{\cmark}{\ding{51}}%
\newcommand{\xmark}{\ding{55}}%
\usepackage{graphicx}  
\usepackage{dcolumn}  
\usepackage{amsmath}
\usepackage{multirow}
\usepackage{bm}   
 \usepackage[colorlinks=true,linkcolor=blue]{hyperref}
\usepackage{amssymb}   
\usepackage{lipsum}
\usepackage{tablefootnote}
\usepackage{hhline}
\usepackage{booktabs}
\usepackage{tabularx}
\usepackage{arydshln}
\usepackage{xifthen}
\usepackage{float}
\usepackage{pifont}
\usepackage{fancyhdr}

\begin{document}
	\title{\textsc{El\textit{A}Tools}: A tool for analyzing anisotropic elastic properties of the 2D and 3D materials}
	\author{Shahram Yalameha}
	\email{yalameha93@gmail.com; sh.yalameha@sci.ui.ac.ir}
 	\affiliation{Faculty of Physics,  University of Isfahan, 81746-73441, Isfahan, Iran.}
	\author{Zahra Nourbakhsh}
	\affiliation{Faculty of Physics,  University of Isfahan, 81746-73441, Isfahan, Iran.}
		\author{Daryoosh Vashaee }
    	\email{dvashae@ncsu.edu}
       \affiliation{Department of Electrical and Computer Engineering, North Carolina State University, Raleigh, NC 27606, USA;}
      \affiliation{Department of Materials Science and Engineering, North Carolina State University, Raleigh, NC 27606, USA.}
     	
	\begin{abstract}
		We introduce a computational method and a user-friendly code with a terminal-based graphical user interface (GUI), named \textsc{\textsc{El\textit{A}Tools}}, developed to analyze mechanical and anisotropic elastic properties. \textsc{\textsc{El\textit{A}Tools}} enables facile analysis of the second-order elastic stiffness tensor of two-dimensional (2D) and three-dimensional (3D) crystal systems. It computes and displays the main mechanical properties including the bulk modulus, Young’s modulus, shear modulus, hardness, p-wave modulus, universal anisotropy index, Chung-Buessem anisotropy index, log-Euclidean anisotropy parameter, Cauchy pressures, Poisson's ratio, and Pugh's ratio, using three averaging schemes of Voigt, Reuss, and Hill. It includes an online and offline database from the Materials Project with more than 13,000 elastic stiffness constants for 3D materials. The program supports output files of the well-known computational codes IRelast, IRelast2D, ElaStic, and AELAS. Four types of plotting and visualization tools are integrated to conveniently interface with G{\scriptsize NUPLOT},  X{\scriptsize MGRACE}, view3dscene and plotly libraries, offering immediate post-processing of the results. \textsc{\textsc{El\textit{A}Tools}} provides reliable means to investigate the mechanical stability based on the calculation of six (three) eigenvalues of the elastic tensor in 3D (2D) materials. It can efficiently identify anomalous mechanical properties, such as negative linear compressibility, negative Poisson’s ratio, and highly-anisotropic elastic modulus in 2D and 3D materials, which are central properties to design and develop high-performance nanoscale electromechanical devices. Moreover, \textsc{\textsc{El\textit{A}Tools}} can predict the behavior of the sound velocities and their anisotropic properties, such as acoustic phase/group velocities and power flow angles in materials, by solving the \textit{Christoffel} equation. Six case studies on selected material systems, namely, ZnAu$_2$(CN)$_4$, CrB$_2$, $\delta$-phosphorene, Pd$_2$O$_6$Se$_2$ monolayer, and GaAs, and a hypothetical set of systems with cubic symmetry are presented to demonstrate the descriptive and predictive capabilities of \textsc{\textsc{El\textit{A}Tools}}.
	\end{abstract}
	
	\maketitle
	
	\textbf{}
	
	\textbf{}
	
	\textbf{}
	\\\textbf{Program summary}
\\ \textit{\textbf{Title}}: \textsc{\textsc{El\textit{A}Tools}}
	\\ \textit{\textbf{Licensing provisions}}: GNU General Public Licence 3.0
	\\ \textit{\textbf{Nature of the problem}}: Identifying anisotropic elastic properties of 2D and 3D materials, and calculating acoustic phase and group velocities in homogeneous solids.
	\\ \textit{\textbf{Solution method}}: Second-order elastic stiffness tensor analysis using transformation law and calculations of the elastic surfaces properties. Solving the \textit{Christoffel} equation eigenvalue problem using diagonalization and calculations of the sound velocities.
	\\  \textit{\textbf{Programming language}}: Fortran 90
	\\ \textit{\textbf{Operating system}}: Unix/Linux/MacOS/Windows by Cygwin: http://www.cygwin.com/
	\\ \textit{\textbf{Distribution format}}: tar.gz
	\\\textit{\textbf{Required routines/libraries}}:  \textsc{LaPack}, \textsc{Blas}, and Plotly Javascript libraries,  G{\footnotesize NUPLOT}, X{\footnotesize MGRACE}, view3dscene.
	\\ \textit{\textbf{Computer}}: Any system with a Fortran 90 (F90) compiler
	\\ \textit{\textbf{Memory}}: Up to 1 GB for any symmetry
	\\ \textit{\textbf{Run time}}: Up to 70 seconds for any symmetry, and (400$\times$400) = ($\theta$$\times$$\phi$)-mesh in  spherical coordinate
	\\\textit{\textbf{Documentation}}: Available at  \href{https://yalameha.gitlab.io/elastictools/index.html}{https://yalameha.gitlab.io/elastictools/index.html}

\section{INTRODUCTION}\label{section:1}
The second-order elastic constants are essential materials parameters, playing pivotal roles in many research areas of engineering \cite{R1, R2}, medical \cite{R3,R4}, condensed matter physics ~\cite{R5,R6, R7}, materials science \cite{R8}, geophysics \cite{R9}, and chemical \cite{R10}. Moreover, the second-order elastic constants will contain information about how acoustic waves behave \cite{R54}. Despite these critical practical features, the second-order elastic constants have been measured for a tiny fraction of known crystalline materials. The small data availability is due to the unavailability of large single crystals for many materials and the difficulty of precise experimental measurements \cite{R11}. The lack of such experimental data limits the scientists' ability to design and develop novel materials. With the development of high-performance computing resources and density functional theory (DFT) \cite{R12}, the determination of elastic constants of many materials can become a reality. DFT is a robust technique able to solve many-body problems using Kohn-Sham equations \cite{R13,R14}, based on which several quantum-chemistry and solid-state physics software have been developed. A number of packages have been introduced to calculate the second-order elastic stiffness tensor of 2D and 3D crystals. For example, {\small VASP} \cite{R15} can calculate second-order elastic constants using strain--stress relationships, and {\small CRYSTAL14} \cite{R16} can compute the piezoelectric and the photoelastic tensors. There is also some software capable of calculating the elastic constants using internal or external packages. For example, {\small WIEN2k} \cite{R17} uses internal packages such as IRelast \cite{R18} and Elast (for cubic systems) to calculate the second-order elastic constants, while \textsf{ElaStic} code is an external package \cite{R19} used in {\small Quantum Espresso} \cite{R53}, Exciting \cite{R87}, and {\small WIEN2k}. With the availability of these packages and the creation of elastic constant databases \cite{R20}, tools for their analysis and visualization have become more significantly desired than ever.
	
\begin{figure}
	\includegraphics[scale=1.0]{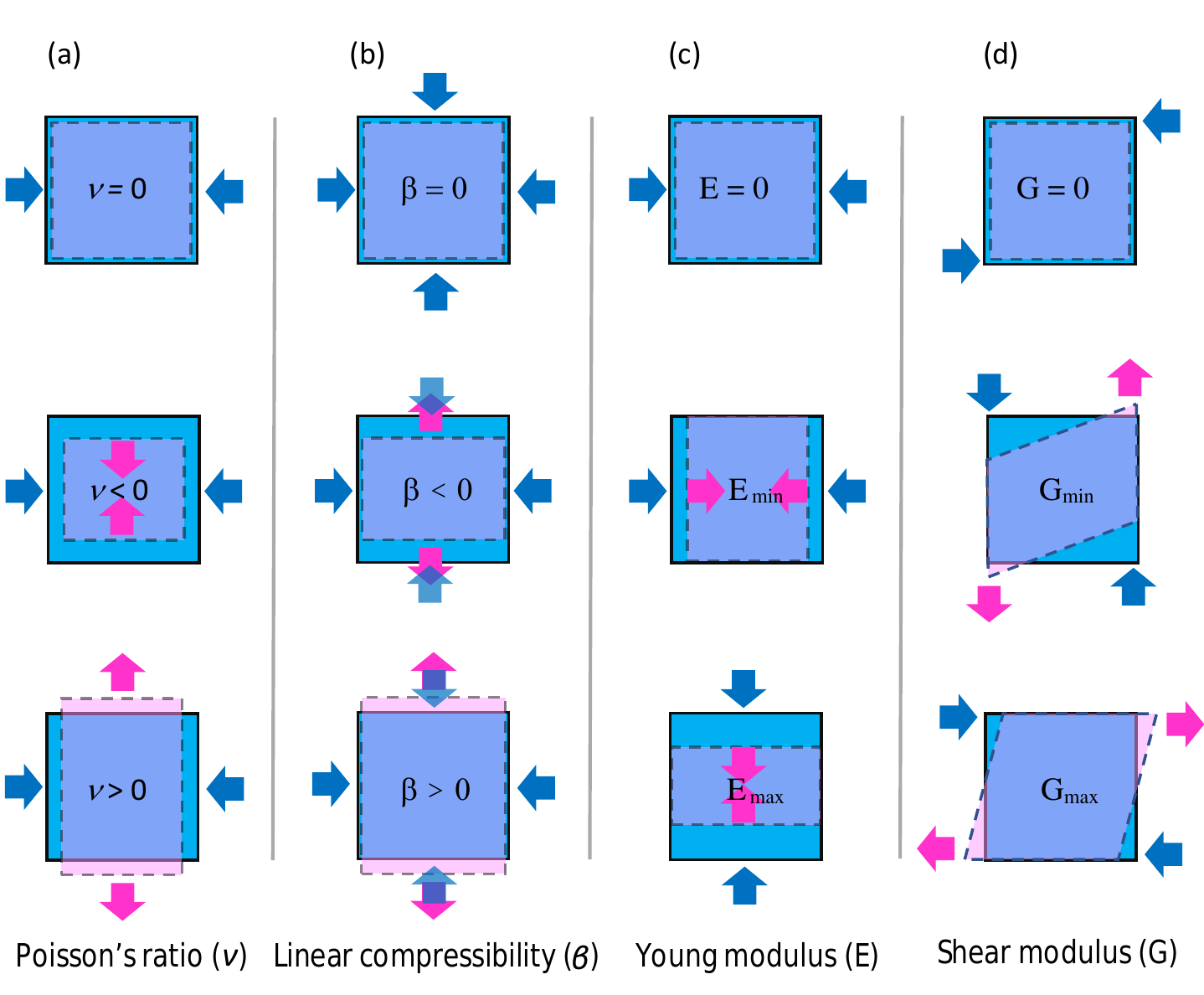} 
	\caption{\label{fig:wide_1}Schematic representation of the directional (a) Poisson's ratio ($\nu$), (b) linear compressibility ($\beta$), (c) shear modulus (\textit{G}), and (d) Young’s modulus (\textit{E}). Blue arrows represent the direction of the stress exerted, and pink arrows the axis show along which the response is measured.}
\end{figure}

In the last two decades, owing to the observation of anomalous mechanical properties in some materials, much effort has been taken to discover and investigate materials with such features. The negative linear compressibility (NLC) \cite{R21}, negative Poisson's ratio (NPR) (\textit{auxetic} material), \cite{R22, R23} and highly-anisotropic elastic modulus \cite{R24} are the most critical anomalous elastic properties that appear in some materials due to stress and strain. These characteristics are visible by analysis and visualization of elastic tensors. When a material is unusually loaded in tension, it extends in the direction of the applied load, and a lateral deformation accompanies its extension. These lateral deformations are quantified by a mechanical property known as the Poisson’s ratio. Poisson’s ratio is defined as the ratio of the negative values of lateral/transverse strain to the longitudinal strain under uniaxial stress. In a material with a positive Poisson’s ratio, when compressive (tensile) stress is acting in one direction, the material tends to expand (shrink) in the perpendicular direction. However, materials with the NPR show opposite behavior  (Fig.~\ref{fig:wide_1}(a)). This feature was considered in 1998 \cite{R30}, although, in 1987, NPR was firstly produced by Lakes \cite{R25} from conventional low-density open-cell polymer foams. In recent years, there have been increasing interests in exploring the possibility of the \textit{auxetic} phenomenon in 2D and 3D materials to design and develop high-performance nanoscale electromechanical devices.

In addition to NPR, another unusual elastic property called NLC \cite{R26, R27} which is resulted from applying hydrostatic pressure to 3D materials leading to an expansion in one direction (Fig.~\ref{fig:wide_1}(b)), has been observed in some materials. The NLC was firstly reported in tellurium in 1922 \cite{R28}. The recent discoveries suggest that NLC is not as rare as previously considered and many materials can offer such a property \cite{R26, R29}. Currently, two software packages and a Python library are available to analyze second-order elastic tensors and visualize elastic properties of 3D materials that can investigate such properties. The first code is {\small ELAM} developed by Marmier \cite{R30}. {\small ElAM}, implemented in Fortran90, is command-line driven and can output 2D cut figures in PostScript (PS) format and 3D surfaces in the Virtual Reality Modelling Language format (VRML). The second one, which was developed by R. Gaillac \textit{et al. }\cite{R31} is {\small ELATE}. {\small ELATE} is a Python module for manipulating elastic tensors and a standalone online application for routine analysis of elastic tensors. In this code, a Python module is used to generate the HTML web page with embedded Javascript for dynamical plots. Notably, this code can import elastic data directly by using the Materials API \cite{R20}. The  \textsc{MechElastic} Python library was also developed by Sobhit Singh et al. \cite{R55}. In this library,  the {\small ELATE} has been used as a module to analyze the properties of elastic anisotropy and auxetic features in which allows direct visualization of the 3D spherical plot, as well as 2D projections on the XY, XZ and YZ planes. \textsc{MechElastic}, powered by the {\small ELATE} module in addition to these features, can calculate changes in compressive and shear velocities, velocity ratios, and Debye velocity estimates by adding mass density. Further, this package can plot the equation of state (EOS) curves for energy and pressure for a variety of EOS models such as Murnaghan, Birch, Birch-Murnaghan, and Vinet, by reading the inputted energy/pressure versus volume data obtained via numerical calculations or experiments. The \texttt{matplotlib} \cite{R56} and \texttt{pyvista} \cite{R57} packages are used to visualize 2D figures and 3D surfaces in the \textsc{MechElastic}.

The present work’s primary motivation is to introduce a comprehensive and efficient program that accommodates all the features of these codes in one place, add some new features, and address their shortcomings. Table \ref{Tab:0} compares \textsc{El\textit{A}Tools} features with \textsc{ElAM}, \textsc{ELATE}, and  \textsc{MechElastic}. Note that other new features may be added in the updates of these packages in the future.
\\\\
 
\begin{sidewaystable} 
	\def\checkmark{\tikz\fill[scale=0.35](0,.39) -- (.25,0) -- (1,.8) -- (.25,.15) -- cycle;}
	\centering
	\caption{Comparison of \textsc{El\textit{A}Tools} with the available primary tools for analyzing anisotropic elastic properties.}
	\label{Tab:0} 
 \centering
	\begin{tabular}{cccccc}  
		
		\hhline{======} 
		\multicolumn{2}{c}{\textbf{Features}}                                                                                                                                                                                               & \multicolumn{1}{l}{\textbf{MechElastic}} & \multicolumn{1}{l}{\textbf{ELAM}} & \multicolumn{1}{l}{\textbf{ELATE}} & \multicolumn{1}{l}{\textbf{ElATools}}  \\ 
		\hhline{======}
		\multicolumn{2}{c}{Main mechanical properties in 3D (see \textbf{Appendix A} )}                                                                                                                                                                                   & \cmark                                                                      & \cmark                                                               & \cmark                                                                & \cmark                                                                     \\ 
		\hhline{------}
		\multicolumn{2}{c}{Main mechanical properties in 2D}                                                                                                                                                                                                                        & \cmark                                                                      & \xmark                                                               & \xmark                                                                & \cmark                                                                    \\ 	
		\hhline{------}	
		\multicolumn{2}{c}{Hardness information}                                                                                                                                                                                                                        & \cmark                                                                      & \xmark                                                               & \xmark                                                                & \cmark                                                                    \\ 
		\hhline{------}
		\multicolumn{2}{c}{Elastic tensor eigenvalues in 3D}                                                                                                                                                                           & \cmark                                                                      & \xmark                                                               & \cmark                                                                & \cmark                                                                   \\ 
		\hhline{------}
		\multicolumn{2}{c}{Elastic tensor eigenvalues in 2D}                                                                                                                                                                              &  \cmark                                                                    & \xmark                                                             & \xmark                                                               & \cmark                                                                    \\ 
		\hhline{------}
		\multirow{8}{*}{Visualization of the 3D surfaces in 3D materials}                                                                          & Shear modulus             & \cmark                                                                      & \cmark                                                               & \cmark                                                                & \cmark                                                                    \\ 
		\cline{2-6}
		& Poisson's ratio           & \cmark                                                                      & \cmark                                                               & \cmark                                                                & \cmark                                                                    \\ 
	    \cline{2-6}
		& Pugh ratio           & \cmark                                                                      & \xmark                                                               & \xmark                                                                & \cmark                                                                     \\
		\cline{2-6}
		& Linear compressibility    & \cmark                                                                      & \cmark                                                               & \cmark                                                                & \cmark                                                                    \\ 
		\cline{2-6}
		& Bulk modulus              & \cmark                                                                    & \xmark                                                               & \xmark                                                                & \cmark                                                                    \\ 
		\cline{2-6}
		& Young's modulus            & \cmark                                                                      & \cmark                                                               & \cmark                                                                & \cmark                                                                    \\ 
		\cline{2-6}
		& Phase velocities                                        & \xmark                                                                      & \xmark                                                               & \xmark                                                               & \cmark                                                                    \\ 
		\cline{2-6}
		& Group velocities                                       & \xmark                                                                      & \xmark                                                               & \xmark                                                                & \cmark                                                                    \\ 
		\cline{2-6}
		& Power flow angles                                       & \xmark                                                                      & \xmark                                                               & \xmark                                                                & \cmark                                                                    \\
		\cline{2-6} 
		& Minimum thermal conductivity                                       & \xmark                                                                      & \xmark                                                               & \xmark                                                                & \cmark                                                                    \\ 		
		\hhline{------}
		\multicolumn{2}{c}{\begin{tabular}[c]{@{}c@{}}Visualization of the 2D projections on an arbitrary plane in 3D materials\end{tabular}}   & \xmark                                                                      & \xmark                                                               & \xmark                                                                & \cmark                                                                    \\ 
		\hhline{------}
		\multirow{3}{*}{Visualization of 2D polar covers in 2D materials}                                                                        & Shear modulus            & 
		\xmark                                                                      & \xmark                                                               & \xmark                                                                & \cmark                                                                    \\
		\cline{2-6}
		& Poisson's ratio           & 	\xmark                                                                      & \xmark                                                               & \xmark                                                                & \cmark                                                                    \\
		\cline{2-6}
		& Young's modulus           & 	\xmark                                                                      & \xmark                                                               & \xmark                                                                & \cmark                                                                    \\
		\hhline{------}
		\multicolumn{2}{c}{\begin{tabular}[c]{@{}c@{}}Visualization of the 2D heat-map and polar heat-map in 3D and 2D~materials\end{tabular}}                                                              & 	\xmark                                                                      & \xmark                                                               & \xmark                                                                & \cmark                                                                    \\
		\hhline{------}		
		\multirow{2}{*}{\begin{tabular}[c]{@{}c@{}}The database of elastic tensors \\(Materials Project’s database)\end{tabular}}                                                                              & Offline                                                & \xmark                                                                      & \xmark                                                               & \xmark                                                                & \cmark                                                                    \\ 
		\cline{2-6}
		& Online                                               & \cmark                                                                      & \xmark                                                               & \cmark                                                                & \cmark                                                                    \\ 
		\hhline{------}
		\multirow{5}{*}{\begin{tabular}[c]{@{}c@{}}Various output formats for custom drawing\\or displaying with different software\end{tabular}} & VRLM format                                             &
		 \xmark                                                                      & \cmark                                                               & \xmark                                                                & \cmark                                                                    \\ 
		\cline{2-6}
		& HTML (offline) format                                    & \xmark                                                                      & \xmark                                                               & \xmark                                                                & \cmark                                                                    \\ 
		\cline{2-6}
		& agr format                                             & \xmark                                                                      & \xmark                                                               & \xmark                                                                & \cmark                                                                    \\ 
		\cline{2-6}
		& gpi format                                              & \xmark                                                                      & \xmark                                                               & \xmark                                                                & \cmark                                                                    \\ 
		\cline{2-6}
		& dat format                                              & \xmark                                                                      & \xmark                                                               & \xmark                                                                & \cmark                                                                    \\
		\hhline{======}
	\end{tabular}
  
\end{sidewaystable}

	The highlighted features of \textsc{El\textit{A}Tools} are as follows:
	
	\begin{itemize}
		\item  Compute and display the main mechanical properties such as Young’s modulus, shear modulus, p-wave modulus, universal anisotropy index \cite{R32}, Chung-Buessem anisotropy index \cite{R32}, log-Euclidean anisotropy parameter \cite{R33}, Kleinman's parameter \cite{R58}, Hardness information, Cauchy pressure, Poisson’s ratio, Pugh’s ratio, according to the three averaging schemes of Voigt, Reuss, and Hill \cite{R34} - Many of these features are included in {\small ELAM}, {\small ELATE}, and  \textsc{MechElastic} codes.
		
		\item  Investigation of mechanical stability using calculation of six (three) eigenvalues of the elastic tensor in 3D (2D) materials - This option exists only in {\small ELATE} (for 3D materials) and \textsc{MechElastic} (for 2D and 3D materials).
		
		\item  Visualization of the 3D surfaces and 2D projections on any desired plane for shear modulus, Poisson’s ratio, Pugh ratio, linear compressibility, bulk modulus, and Young’s modulus. - {\small ELATE} and \textsc{MechElastic} only depict these features on XY, XZ, and ZY planes. Currently, visualization of the 3D surfaces and 2D projections on any desired plane for bulk modulus and Pugh ratio does not exist in either {\small ELAM} or {\small ELATE}.
		
		\item  Visualization of the 3D surfaces and 2D projections on any desired plane for phase velocities, group velocities and power flow angles (includes primary,  fast-secondary and slow-secondary modes) in 3D materials - These options do not exist in the {\small ELAM} and {\small ELATE} or \textsc{MechElastic}.
		
		\item  Visualization of the 2D polar covers for Poisson’s ratio, shear modulus, and Young’s modulus in 2D materials - This option does not exist in the {\small ELATE}, \textsc{MechElastic}, and {\small ELAM}.
		
		\item  An offline/online database of more than 13000 elastic tensors taken from Materials API (Materials Project database) - This option is also available online in {\small ELATE} and \textsc{MechElastic}.
		
		\item  Supports various output formats for custom drawing or displaying with different software: ``.dat'' files for standard plotting, ``.wrl'' files for visualization of the 3D surfaces by view3dscene software \cite{R35}, ``.agr'' files for visualization of the 2D projections that are opened by X{\scriptsize MGRACE} \cite{R36}, ``.html'' files for visualization of the 3D surfaces by any Web browser, and ``.gpi'' script files for visualization of the 3D surfaces, 2D heat maps (for 3D materials), polar-heat maps (for 2D materials), and 2D projections that can be run by G{\footnotesize NUPLOT} \cite{R37} - {\small ELAM} generates only VRML (for 3D visualization) and PS formats (for 2D cut visualization). {\small ELATE} provides images online in PNG format only. \textsc{MechElastic} displays properties by \texttt{matplotlib} and \texttt{pyvista} python libraries.
		\item A user-friendly code with a terminal-based graphical user interface (GUI). A summary of this terminal-based GUI is included in \textbf{Supplementary Information}. - This feature does not exist in the {\small ELAM} and {\small ELATE} or \textsc{MechElastic}.
	\end{itemize}

The rest of the paper is arranged as follows: The theoretical background of elasticity and analysis of elastic tensor is explained in detail in the next section. Package descriptions, including the workflow, code structure, installation, input and output files, visualization data, and test cases are presented in Sec. 3. Sec. 4 is summary and outlook. Finally, appendices A, B, C, D and Supplementary Information file are provided as complementary parts.
\section{ Theoretical Background} \label{section:2}
\subsection{Hooke's law and elastic tensor of crystals }
\begin{figure}
	\includegraphics[scale=1]{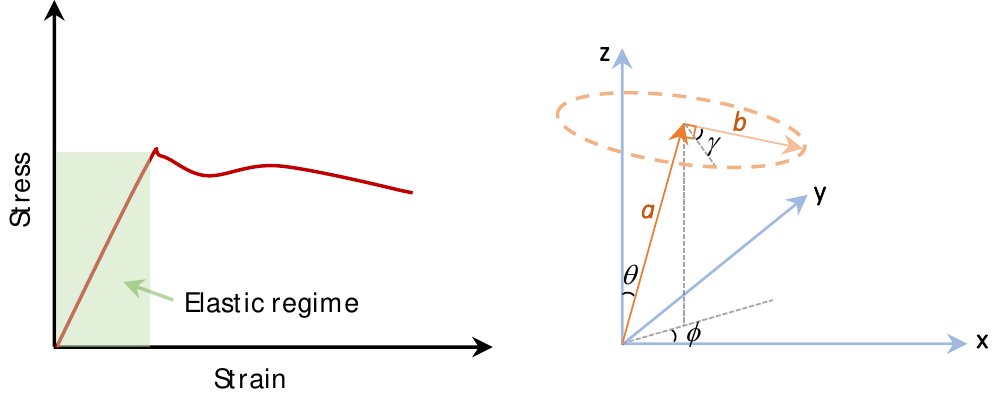} 
	\caption{\label{fig:wide_2}Schematic stress–strain curves and definitions of angles used to describe the directions in the calculations related to \textsc{El\textit{A}Tools}. The elastic regime corresponds to the portion of the diagram where the strain is proportional to the stress.}
\end{figure}

The shape of solid material changes when subjected to stress. Provided that the stress is below a specific value, the elastic limit, the strain is recoverable (Fig.~\ref{fig:wide_2}). 
This means that the material returns to its original shape when the stress is removed. In
this elastic regime, according to Hooke’s law, it can be stated that for sufficiently low stresses, the amount of strain is proportional to the magnitude of the applied stress:
\begin{equation} \label{1)} 
\varepsilon =S\sigma , 
\end{equation} 
where \textit{S} is a constant. \textit{S} is called the \textit{elastic compliance constant } (ECC), or the \textit{compliance}. Another form of this equation can be written as follows,
\begin{equation}\label{2)}  
\varepsilon =C\sigma ,\, \, \, E\equiv C=1/s,    
\end{equation} 
where \textit{C} is the \textit{elastic stiffness constant} (ESC), or \textit{stiffness}, and E is Young’s modulus. The general form of Hooke's law can be rewritten in the form of tensor,
\begin{equation} \label{3)} 
\varepsilon _{ij} =S_{ijkl} \sigma _{kl} , 
\end{equation}
where \textit{S$_{ijkl}$} are the ECCs of the crystal. Also, as an alternative to Eq.\eqref{2)},
\begin{equation} \label{4)} 
\sigma _{ij} =c_{ijkl} \varepsilon _{kl} , 
\end{equation} 
where \textit{C$_{ijkl}$} are the ESCs. Eq.\eqref{3)} and Eq.\eqref{4)} stand for nine equations, each with nine terms on the right-hand side. The \textit{S$_{ijkl}$} or \textit{C$_{ijkl}$} are 4${}^{th}$-rank tensors, and \textit{$\epsilon$$_{ij}$} or \textit{$\sigma$$_{ij}$}$_{\ }$ are  second\textit{${}^{th}$}-rank tensors. Hence, the \textit{C$_{ijkl\ }$}(\textit{S$_{ijkl}$}) consists of 81 \textit{stiffness} (\textit{compliances}) constants of the crystal. Due to the inherent symmetries (translational and rotational symmetries) of $\epsilon$$_{ij}$, $\sigma$$_{ij}$, and \textit{S$_{ijkl}$} or \textit{C$_{ijkl,\ }$} the number of independent coordinates of the 4\textit{${}^{th}$}-rank tensor reduces to 21 for the least symmetric case. On the other hand, the further reduction resulting from the symmetry of the crystal can be applied to this number: 21 for triclinic, 15 for monoclinic, 9 for orthorhombic, 7 for trigonal, 5 for tetragonal, 5 for hexagonal and 3 for cubic.
\subsection{Transformation law, Christoffel equation, and representation surfaces of elastic properties }
A 4\textit{${}^{th}$-}rank tensor is defined (like tensors of lower rank) by its transformation law \cite{R38}. We know that the 81 tensor components \textit{$A_{\, ijkl}$} representing a physical quantity are said to form a 4${}^{th}$-rank tensor if they transform on change of axes to \textit{$A'_{\, ijkl}$}, where
\begin{equation} \label{5)} 
A'_{\, ijkl} =a_{im} \, a_{jn} \, a_{ko} \, a_{lp} A_{mnop} . 
\end{equation} 
It can be shown that 4\textit{${}^{th}$}-rank tensor \textit{S$_{ijkl}$} or \textit{C$_{ijkl}$} follows this rule \cite{R38}:
\begin{equation} \label{6)} 
\left. \begin{array}{l} {\varepsilon '_{ij} =a_{ik} a_{jl} \varepsilon _{kl} ,} \\ {\varepsilon _{kl} =S_{klmn} \sigma _{mn} ,} \\ {\sigma _{mn} =a_{om} a_{pn} \sigma '_{op} ,} \end{array}\right\}\varepsilon '_{ij} =a_{ik} a_{jl} \, S_{klmn} \, a_{om} a_{pn} \, \sigma '_{op} . 
\end{equation} 
By comparing Eq.\eqref{3)} and the recent equation, we have: 
\begin{equation} \label{7)} 
S'_{ijkl} =a_{im} a_{jn} a_{ko} a_{lp} S_{mnop} \, , 
\end{equation} 
which is the necessary transformation law. To express the anisotropic form of Hooke's law in matrix notation, we use the Voigt notation scheme. In the \textit{$S'_{ijkl}$} and \textit{S$_{mnop}$}, the first two suffixes are abbreviated into a single one running from 1 to 6, and the last two are abbreviated in the same way, according to the following Voigt scheme:
\begin{equation} \label{8)} 
\begin{array}{l} {{\rm Tensor\; notation}:{\rm \; }11{\rm \; }\, \, 22\, \, {\rm \; }33{\rm \; }\, \, 23,32\, \, \, {\rm \; }31,13\, \, \, {\rm \; }12,21} \\ {{\rm Matrix\; notation}:{\rm \; }\, 1{\rm \; }\, \, \, \, \, 2{\rm \; }\, {\rm \; }\, \, \, 3{\rm \; }\, \, \, \, \, \, \, 4{\rm \; }\, \, \, \, \, \, \, \, \, \, \, \, \, 5{\rm \; }\, \, \, \, \, \, \, \, \, \, \, 6} \end{array} 
\end{equation} 
Therefore, the components of the stress (\textit{$\sigma$}) and the strain (\textit{$\epsilon$}) tensors are written in a single suffix running from 1 to 6,
    \begin{equation} \label{9)} 
\begin{array}{l} {\sigma _{ij} =\left(\begin{array}{ccc} {\sigma _{11} } & {\sigma _{12} } & {\sigma _{13} } \\ {\sigma _{12} } & {\sigma _{22} } & {\sigma _{23} } \\ {\sigma _{13} } & {\sigma _{23} } & {\sigma _{33} } \end{array}\right)\stackrel{{\rm Voigt}\, \, {\rm scheme}}{\longrightarrow}\left(\begin{array}{ccc} {\sigma _{1} } & {\sigma _{6} } & {\sigma _{5} } \\ {\sigma _{6} } & {\sigma _{2} } & {\sigma _{4} } \\ {\sigma _{5} } & {\sigma _{4} } & {\sigma _{3} } \end{array}\right),} \\\\ {\varepsilon _{ij} =\left(\begin{array}{ccc} {\varepsilon _{11} } & {\varepsilon _{12} } & {\varepsilon _{13} } \\ {\varepsilon _{12} } & {\varepsilon _{22} } & {\varepsilon _{23} } \\ {\varepsilon _{13} } & {\varepsilon _{23} } & {\varepsilon _{33} } \end{array}\right)\stackrel{{\rm Voigt}\, \, {\rm scheme}}{\longrightarrow}\left(\begin{array}{ccc} {\varepsilon _{1} } & {{\textstyle\frac{1}{2}} \varepsilon _{6} } & {{\textstyle\frac{1}{2}} \varepsilon _{5} } \\ {{\textstyle\frac{1}{2}} \varepsilon _{6} } & {\varepsilon _{2} } & {{\textstyle\frac{1}{2}} \varepsilon _{4} } \\ {{\textstyle\frac{1}{2}} \varepsilon _{5} } & {{\textstyle\frac{1}{2}} \varepsilon _{4} } & {\varepsilon _{3} } \end{array}\right).} \end{array} 
\end{equation} 
According to this scheme, we have for the \textit{S$_{mnop}$$_{\ }$}\cite{R38},
\begin{center}

	\noindent \textit{S$_{mnop\ }$}=\textit{ S$_{ij}$}, when \textit{i} and \textit{j} are 1; 2 or 3,
	
	\noindent 2\textit{S$_{mnop}$} = \textit{S$_{ij}$}, when either \textit{i} or \textit{j }are 4; 5 or 6,
	
	\noindent 4\textit{S$_{mnop}$} = \textit{S$_{ij\ }$}when both \textit{i }and \textit{j} are 4; 5; or 6.
\end{center}
\noindent Therefore, Eq.\eqref{3)} takes the shorter form:
\begin{equation} \label{10)} 
\varepsilon _{i} =S_{ij} \sigma _{j} \, \, \, \, (i,j=1,{\rm \; }2,...,{\rm \; }6). 
\end{equation} 
The Voigt scheme replaces the cumbersome 2${}^{th}$ and 4\textit{${}^{th}$}-rank tensors in a 3-dimensional vector space of vectors and matrices in a 6-dimensional vector space. The reason for introducing the factors of 0.5 in Eq.\eqref{9)} and the factors of 2 and 4 into the definitions of the \textit{S$_{ij}$} is to enable writing Eq.\eqref{10)} in a compact form.

\begin{figure}
	\includegraphics[scale=1.1]{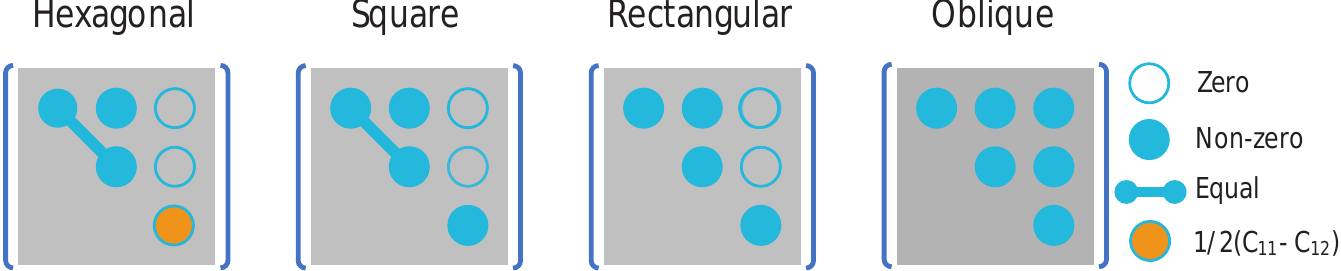} 
	\caption{\label{fig:wide_31}Classification of crystal systems and independent elastic constants for 2D materials.}
\end{figure}

Using \textit{S$_{ij}$} in Eq.\eqref{10)} and Eq.\eqref{5)}, we can get a general and and straightforward compliance transformation relation for any crystal from the old systems (${T'}$) to measurement systems (\textit{T)}:
\begin{equation} \label{11)} 
T'_{\, ijkl} =r_{i\alpha } \, r_{j\beta } \, r_{k\gamma } \, r_{l\delta } T_{\alpha \beta \gamma \delta } , 
\end{equation} 
where the \textit{r} represents the components of the rotation matrix (or direction cosines). In general, the tension produces not only longitudinal, and lateral strains, but shear strains as well. Therefore, spherical coordinates are suitable for such stresses and the responses that materials give to stresses. We choose \textbf{\textit{r}}$\mathrm{\equiv}$\textbf{\textit{a}} to be the first unit vector in the new basis set \cite{R30,R31,R50},
\begin{equation} \label{12)} 
\textbf{\textit{a}}=\left(\begin{array}{c} {\sin (\theta )\cos (\varphi )} \\ {\sin (\theta )\sin (\varphi )} \\ {\cos (\theta )} \end{array}\right);\, \, \, \, \, \, \, 0\, \le \theta \le \pi ,\, \, \, 0\le \varphi \le \pi , 
\end{equation} 
This unit vector (\textbf{\textit{a}}) is required to determine Young's modulus (\textit{E}), linear compressibility (\textit{$\beta$}), and bulk modulus (\textit{B}). But some elastic properties such as the shear modulus (\textit{G}) and Poisson's ratio (\textit{v}) requires another perpendicular direction. Therefore, we define unit vector \textbf{\textit{b}}, which is perpendicular to \textbf{\textit{a }}(see Fig.~\ref{fig:wide_2}), as follows \cite{R30,R31}:
\begin{equation} \label{13)} 
\textbf{\textit{b}}\equiv \left(\begin{array}{c} {\cos (\theta )\cos (\varphi )\cos (\gamma )-\sin (\theta )\sin (\gamma )} \\ {\cos (\theta )\sin (\varphi )\cos (\gamma )-\cos (\theta )\sin (\gamma )} \\ {-\sin (\theta )\cos (\gamma )} \end{array}\right),\, \, \, \, 0\le \, \gamma \le 2\pi .\, 0  
\end{equation} 
Therefore, by defining these two vectors, Eq.\eqref{11)} is as follows:
\begin{equation} \label{14)} 
T'_{\, \alpha \beta \gamma \delta } =a_{\alpha i} \, a_{\beta j} \, b_{\gamma k} \, b_{\delta l} T_{ijkl} , 
\end{equation} 
Using this equation, we can calculate the representation surfaces for elastic properties. For instance, from Eq.\eqref{10)}, we know that Young's modulus can be obtained by using purely normal stress (see Fig.~\ref{fig:wide_1}(c)),
\begin{equation} \label{15)} 
\begin{array}{l} {E(\textbf{a})\equiv \frac{1}{S'_{1111} } =\sum_{ijkl}^6 \frac{1}{a_{1i} a_{1j} a_{1k} a_{1l} S_{ijkl} } } \\\\ \, \, \, \, \, \, \, \, \, \, \,\, \, \, \, \,\, \, \, \, \, \, \, \, \, \, \,\, \,\, \,\, \ =\frac{1}{S'_{11} } =\frac{1} {a_{i} a_{j} a_{k} a_{l} S_{ijkl}}  \, \, \, \, ({\rm by}\, {\rm Einstein{'} s\; summation\; rule}), \end{array} 
\end{equation} 
The volume compressibility of a crystal is the proportional decrease in volume of a crystal when subjected to unit hydrostatic pressure, but the linear compressibility is the relative decrease in length of a line when the crystal is subjected to unit hydrostatic pressure (HP). Hence, it is obtained by applying isotropic stress (P$_{HP}$) in a tensor form so that \textit{$\varepsilon$$_{ij}$}=-(P$_{HP}$)\textit{S$_{ijkk}$} \cite{R30,R38}, and by considering that the extension in \textit{a}  direction is \textit{$\varepsilon$$_{ij}$a$_{i}$a$_{j}$}, and for this reason, we have:
\begin{equation} \label{16)} 
\beta (\textbf{\textit{a}})=a_{i} a_{j} a_{k} a_{k} S_{ijkk} =a_{i} a_{j} S_{ijkk} \, ;\, \, \, a_{k} .\, a_{k} =1 
\end{equation} 
On the other hand, the relationship between \textit{$\beta$} and \textit{B} can be expressed as follows \cite{R38},
\begin{equation} \label{17)} 
B(\textbf{\textit{a}})=\frac{1}{\beta (\textbf{a})} =\frac{1}{a_{i} a_{j} S_{ijkk} }  
\end{equation} 
As mentioned, the \textit{G} and \textit{$\nu$} are not as straightforward to represent and depend on two directions (\textit{\textbf{a}} and \textit{\textbf{b}}). The shear ratio (Poisson's ratio) is obtained by applying a pure shear (Fig.~\ref{fig:wide_1}(d))  (a purely normal) stress in the vector form of Eq.\eqref{3)}, and results in:
\begin{equation} \label{18)} 
G(\textbf{\textit{a}},\textbf{\textit{b}})=\frac{1}{4S_{1212} } =\frac{1}{4S_{66} } =\frac{1}{4} \frac{1}{a_{i} b_{j} a_{k} b_{l} S_{ijkl} } , 
\end{equation} 
\begin{equation} \label{19)} 
\nu (\textbf{\textit{a}},\textbf{\textit{b}})=-\frac{S'_{1212} }{S'_{1111} } =-\frac{S'_{12} }{S'_{11} } =-\frac{a_{i} a_{j} b_{k} b_{l} S_{ijkl} }{a_{i} a_{j} a_{k} a_{l} S_{ijkl} } . 
\end{equation} 
To better understand these equations, we obtain Young's modulus of a cubic crystal. As for cubic crystal, because of lattice symmetry, there are three independent variables \textit{C}$_{11}$, \textit{C}$_{12}$, \textit{C}$_{44}$ in the \textit{C$_{ij}$}, and \textit{S}$_{11}$, \textit{S}$_{12}$$_{,}$ and \textit{S}$_{33}$ in the \textit{S$_{ij}$}. Using Eq.\eqref {15)} and Eq.\eqref {12)}, we have:
\begin{equation} \label{20)} 
\begin{array}{l} {S'_{1111} =a_{11} a_{11} a_{11} a_{11} S_{1111} +a_{12} a_{12} a_{12} a_{12} S_{2222} +a_{13} a_{13} a_{13} a_{13} S_{3333} +} \\\\ {\, \, \, \, \, \, \, \, \, \, \,\,\,\,\, \, \, \,\,\, a_{11} a_{11} a_{12} a_{12} S_{1122} +a_{11} a_{11} a_{13} a_{13} S_{1133} +a_{12} a_{12} a_{13} a_{13} S_{2233} +} \\\\ {\, \, \, \, \, \, \, \, \, \, \, \, \, \, \,\,\,\,\,\,a_{11} a_{11} a_{12} a_{12} S_{2211} +a_{11} a_{11} a_{13} a_{13} S_{3311} +a_{12} a_{12} a_{13} a_{13} S_{3322} +} \\\\ {\, \, \, \, \, \, \, \, \, \, \, \, \,\,\,\,\,\,\, \frac{1}{4} [a_{12} a_{13} a_{12} a_{13} S_{2323} ]+\frac{1}{4} [a_{13} a_{11} a_{13} a_{11} S_{3131} ]+\frac{1}{4} [a_{11} a_{12} a_{11} a_{12} S_{1212} ]} \\\\ {\, \, \, \, \, \, \, \, \,\,\,\,\,\,  =S_{11} (a_{11}^{4} +a_{12}^{4} +a_{13}^{4} )+(\frac{1}{4} S_{44} +2S_{12} )[a_{12}^{2} a_{13}^{2} +a_{11}^{2} a_{13}^{2} +a_{11}^{2} a_{12}^{2} ],} \end{array} 
\end{equation} 
with further simplification,
\begin{equation} \label{21)} 
\frac{1}{E} =S'_{1111} =S_{11} -2(S_{11} -S_{12} -\frac{1}{2} S_{44} )(a_{11}^{2} a_{12}^{2} +a_{11}^{2} a_{13}^{2} +a_{12}^{2} a_{13}^{2} ) 
\end{equation} 
To calculate the orientation-dependent Poisson's ratio, shear modulus and Young's modulus in 2D material, Eq.\eqref{12)} and Eq.\eqref{13)} are used again and changed as follows:
\begin{equation} \label{22)} 
\textbf{a}=\left(\begin{array}{c} {\cos (\varphi )} \\ {\sin (\varphi )} \\ {0} \end{array}\right);\, \, \, \textbf{b}=\left(\begin{array}{c} {-\sin (\varphi )} \\ {\cos (\varphi )} \\ {0} \end{array}\right)\, ;\, \, \, \, \, \, \, \, 0\le \varphi \le 2\pi .\,  
\end{equation} 
For 2D system, according to Hooke's law (Eq.\eqref{3)} and Eq.\eqref{4)}), the relationship between \textit{$\sigma$} and the corresponding strain tensor \textit{$\varepsilon$ }can be described using the stiffness tensor \textit{C$_{ij}$}, for orthogonal symmetry under plane stress conditions as \cite{R39},
\begin{equation} \label{23)} 
\left(\begin{array}{c} {\sigma _{11} } \\ {\sigma _{22} } \\ {\sigma _{12} } \end{array}\right)=\left(\begin{array}{ccc} {C_{11} } & {C_{12} } & {0} \\ {C_{12} } & {C_{22} } & {0} \\ {0} & {0} & {C_{66} } \end{array}\right)\left(\begin{array}{c} {\varepsilon _{11} } \\ {\varepsilon _{22} } \\ {2\varepsilon _{12} } \end{array}\right), 
\end{equation} 
where Voigt notation has been used for \textit{C$_{ij}$}. Here \textit{C}$_{11\ }$(\textit{S}$_{11}$), \textit{C}$_{22\ }$(\textit{S}$_{22}$), \textit{C}$_{12}$=\textit{C}$_{21\ }$(\textit{S}$_{12}$=\textit{S}$_{21}$), and \textit{C}$_{66\ }$(\textit{S}$_{66}$) represent \textit{C}$_{1111\ }$(\textit{S}$_{1111}$), \textit{C}$_{2222\ }$(\textit{S}$_{2222}$), \textit{C}$_{1122\ }$(\textit{S}$_{1122}$), and \textit{C}$_{1212\ }$(\textit{S}$_{1212}$), respectively. So, using the previous equations, the in-plane Young's modulus, shear modulus, and Poisson's ratio can be defined as:
\begin{equation} \label{24)} 
{E(\varphi )=\frac{1}{S_{11} \cos ^{4} (\varphi )+S_{22} \sin ^{4} (\varphi )+(2S_{12} +S_{66} )\cos ^{2} (\varphi )\sin ^{2} (\varphi )}} , 
\end{equation} 
\begin{equation} \label{25)} 
\nu (\varphi )=-\frac{[S_{11} -S_{66} +S_{22} ]\cos ^{2} (\varphi )\sin ^{2} (\varphi )+S_{12} \cos ^{4} (\varphi )+S_{12} \sin ^{4} (\varphi )}{[2S_{12} +S_{66} ]\cos ^{2} (\varphi )\sin ^{2} (\varphi )+S_{11} \cos ^{4} (\varphi )+S_{22} \sin ^{4} (\varphi )}  ,
\end{equation} 
\begin{equation} \label{26)} 
\frac{1}{4G(\varphi )} =[S_{11} +S_{22} -2S_{12} ]\cos ^{2} (\varphi )\sin ^{2} (\varphi )+\frac{1}{4} S_{66} [\cos ^{4} (\varphi )+\sin ^{4} (\varphi )-2\sin ^{2} (\varphi )\cos ^{2} (\varphi )] .
\end{equation} 
Equations \eqref{24)},\eqref{25)}, and \eqref{26)} can be used for hexagonal, square, and rectangular 2D crystal systems (see Fig.~\ref{fig:wide_31}), which have two, three, and four independent elastic constants, respectively. For two-dimensional oblique systems, there are six independent elastic constants(C$_{16}$ and C$_{26}$ are non-zero). In this case, the above equations are defined as follows~\cite{R59,R60}: 
\begin{equation} \label{27)}
	\centering 
	\begin{split}
	E(\varphi )=\frac{1}{S_{11} \cos ^{4} (\varphi )+S_{22} \sin ^{4} (\varphi )+(2S_{12} +S_{66} )\cos ^{2} (\varphi )\sin ^{2} (\varphi )+}\\
	\frac{1}{2S_{16}\cos ^{3} (\varphi )\sin (\varphi )+2S_{26}\sin ^{3} (\varphi )\cos (\varphi ))} ,
\end{split}
\end{equation} 
\begin{equation} \label{28)} 
	\centering
	\begin{split}
	\nu (\varphi )=-\frac{[S_{22} -S_{66} +S_{22} ]\cos ^{2} (\varphi )\sin ^{2} (\varphi )+S_{12}[\cos ^{4} (\varphi )+\sin ^{4} (\varphi )]+}{[2S_{12} +S_{66} ]\cos ^{2} (\varphi )\sin ^{2} (\varphi )+S_{11} \cos ^{4} (\varphi )+S_{22} \sin ^{4} (\varphi )+} \\
	\frac{S_{16}[\sin ^{3} (\varphi )\cos (\varphi )-\cos ^{3} (\varphi )\sin (\varphi )]+}{2S_{16}\sin^{3}(\varphi )\cos(\varphi )+}\\
	\frac{S_{26}[\cos ^{3} (\varphi )\sin (\varphi )-\sin ^{3} (\varphi )\cos (\varphi )]}{2S_{26}\cos ^{3} (\varphi)\sin(\varphi )},
\end{split}
\end{equation} 
\begin{equation} \label{29)} 
	\centering
\begin{split} {
		\frac{1}{4G(\varphi )} =[S_{11} +S_{22} -2S_{12} ]\cos ^{2} (\varphi )\sin ^{2} (\varphi )+\frac{1}{4} S_{66} [\cos ^{4} (\varphi )+\sin ^{4} (\varphi )-2\sin ^{2} (\varphi )\cos ^{2} (\varphi )]}\\{+ S_{16} [\sin ^{3} (\varphi )\cos  (\varphi ) - \cos ^{3}  (\varphi )\sin (\varphi ) ] + S_{26} [\cos ^{3} (\varphi )\sin  (\varphi ) - \sin ^{3}  (\varphi )\cos (\varphi ) ]}.
\end{split} 
\end{equation} 

 An important relation in material science is the connection between the elastic constants tensor and elastic wave velocities of a solid. Since sound is a form of elastic waves traveling in a homogeneous medium (e.g., a perfect crystal), the C$_{ijkl}$ will contain information about how these sound waves are propagated. Knowing the elastic constants makes it possible to predict the sound velocities in a material by the \textit{Christoffel} equation \cite{R62,R63}, and determining the dispersion relation for these waves is possible by solving this equation:
\begin{equation} \label{30)} 
	[M_{il}-\rho \omega^{2}  \delta_{il} ]s_{l}=0 .
\end{equation}

For a plane wave with wave vector \textbf{q}, frequency $\omega$, and polarization $\hat{s}$ in a material with density $\rho$, the \textit{Christoffel} matrix ($M_{il}$) can be defined as follows:
\begin{equation} \label{31)} 
	M_{il}=q_{j}C_{ijkl}q_{k}.
\end{equation}
Eq.\eqref {30)} is a simple eigenvalue problem that can be routinely solved for arbitrary
\textbf{q}, and the result is a set of three frequencies and polarization vectors for each value of \textbf{q}. The \textit{Christoffel} matrix is symmetric and real; so the eigenvalues are real, and polarization vectors \{$\hat{\textbf{s}}$\}  constitute an orthogonal basis. 

We use the reduced elastic constants tensor $\tilde{C}_{ijkl}$ =  $\rho^{-1}$C$_{ijkl}$ and the corresponding reduced \textit{Christoffel} matrix ($\tilde{M}_{il}$) for further simplification. Due to the wavelength-independence of the velocities, we do not consider \textbf{q} a dimension of inverse length but a dimensionless unit vector that defines only the direction of travel of a  plane wave. This causes the dimension of the $\tilde{M}_{il}$ to change from frequency to velocity squared. Therefore, Eq.\eqref {30)} can be reduced as follows:
\begin{equation} \label{32)} 
	[\tilde{M}_{il}-v^{2}_{p}  \delta_{il} ]s_{l}=0;\,\, v^{2}_{p}=\omega^{2}/q^{2},
\end{equation}
where $v_{p}$ is the velocity of a plane wave traveling in the direction of  \textbf{$\hat{q}$}. The calculations of a material sound velocities based on the Eq.\eqref {32)} are a straightforward eigenvalue problem. From this equation, we obtain three velocities, one primary (P) and two secondary (S), which correspond to the longitudinal and transversal polarizations, respectively. Generally, the velocity of a plane wave is referred to as the \textit{phase velocity} ($v_{p}$). The real sound, like $v_{p}$ is never purely monochromatic nor purely planar. Hence, we consider a wave packet with a small spread in wavelength and direction of travel. The velocity of the wave packet formed by the superposition of these phase waves is called the \textit{group velocity} ($v_{g}$).It is the velocity that acoustic energy travels through a non-dispersive and homogeneous medium and is defined by
\begin{equation} \label{33)} 
  \mathbf{v}_{g} \equiv  \mathbf{\nabla} v_{p},
\end{equation}
where $v_{p}$ is a \textit{scalar} function of \textbf{$\hat{\textbf{q}}$}, and $\textbf{v}_{g}$ is a vector-valued function, which generally does not lie in the direction of \textbf{$\hat{q}$}. The angle between $v_{p}$ and $v_{g}$ is called the \textit{power flow angle} ($\psi$) and is defined as
\begin{equation} \label{34)} 
	 \nu_{p} = \nu_{g} cos(\psi); cos(\psi)=\mathbf{\hat{n}}_{p}. \mathbf{\hat{n}}_{g},
\end{equation}
where $\hat{\textbf{n}}_{p}$ and   $\hat{\textbf{n}}_{g}$ are the normalized directions of the $v_{p}$ and $v_{g}$, respectively.
\subsection{Mechanical properties and elastic anisotropy of 2D and 3D materials}
From elastic constants, other basic elastic properties, including elastic moduli, can be obtained. The elastic response of an isotropic system is generally described by the \textit{B} and the \textit{G}, which may be obtained by averaging the single-crystal elastic constants. The averaging methods most often used are the Voigt \cite{R40}, Reuss \cite{R41} and Hill \cite{R34} bounds. In Voigt's and, Reuss's approximations, the equation takes the following form:
\begin{equation} \label{35)} 
\begin{array}{l} {B_{V} =9^{-1} ([C_{11} +C_{22} +C_{33} ]+2[C_{12} +C_{23} +C_{31} ]),} \\\\ {G_{V} =15^{-1} ([C_{11} +C_{22} +C_{33} ]-[C_{12} +C_{23} +C_{13} ]+3[C_{44} +C_{55} +C_{66} ]),} \\\\ {B_{R} =([S_{11} +S_{22} +S_{33} ]+2[S_{12} +S_{23} +S_{13} ])^{-1},} \\\\ {G_{R} =15 (4[S_{11} +S_{22} +S_{33} ]-4[S_{12} +S_{23} +S_{31} ]+3[S_{44} +S_{55} +S_{66} ])^{-1}.} \end{array} 
\end{equation} 
Also, the arithmetic mean of the Voigt and Reuss bounds, termed the Voigt-Reuss-Hill (VRH) average is also found as a better approximation to the actual elastic behavior of a polycrystal material,
\begin{equation} \label{36)} 
\begin{array}{l} {B_{VRH} =\frac{1}{2} (B_{V} +B_{R} ),} \\\\ {G_{VRH} =\frac{1}{2} (G_{V} +G_{R} )} \end{array} 
\end{equation} 
The Young's modulus (\textit{E}), and Poisson's ratio (\textit{$\nu$}) for an isotropic material are given by:
\begin{equation} \label{37)} 
E=\frac{9BG}{3B+G} ,\, \, \, \, v=\frac{3B-2G}{2(3B+G)} . 
\end{equation} 

    The elastic anisotropy is a crucial measurement of the anisotropy of chemical bonding and can be calculated by elastic constants. For all crystal systems, the bulk response is in general, anisotropic, and one must account for such contributions to quantify the extent of anisotropy accurately. For this purpose, Ranganathan \textit{et al}. \cite{R32} introduce a new universal anisotropy index (\textit{A}${}^{U}$),
\begin{equation} \label{38)} 
A^{U} =5\frac{G_{V} }{G_{R} } +\frac{B_{V} }{B_{R} } -6. 
\end{equation} 

It is noteworthy that in weakly anisotropic materials, i.e. isotropic material, all such averages lead to similar results for elastic moduli.
The mechanical behavior such as \textit{ductile} or \textit{brittle} can be represented by the ratio of the \textit{G} to the \textit{B}, \textit{i.e.,} the Pugh ratio \textit{G}/\textit{B}, by simply considering \textit{B} as the resistance to fracture and \textit{G} as the resistance to plastic deformation. The critical value of the Pugh ratio to separate \textit{ductile} and \textit{brittle} materials is around 0.57. If \textit{G/B }$\mathrm{<}$ 0.57, the material is more \textit{ductile}; otherwise, it behaves in a \textit{brittle} manner \cite{R5,R42}. Hence, a higher Pugh ratio indicates more brittleness property. Cauchy pressure (\textit{P$_{C}$}) is another characteristic to describe the brittleness and ductility of the metals and compounds and  is defined in different symmetries \cite{R43,R51,R52} by:
\begin{equation} \label{39)} 
P_{C}^{c} =C_{12} -C_{44} \, \ (Cubic \, \,symmetry)
\end{equation} 
\begin{equation} \label{40)} 
P_{C}^{a} =C_{13} -C_{44} , P_{C}^{b} =C_{12} -C_{66}  \, \ (Hexagonal,Trigonal, Tetragonal  \, \,symmetries)
\end{equation} 
\begin{equation} \label{41)} 
P_{C}^{a} =C_{23} -C_{44} , P_{C}^{b} =C_{13} -C_{55} , P_{C}^{c} =C_{12} -C_{66} \, \ (Orthorhombic\,\,symmetry)
\end{equation} 
For covalent materials with brittle atomic bonds, the \textit{P$_{C}$} is negative, because in this case, material resistance to shear strain, \textit{i.e.}, \textit{C}$_{44}$, is much more than that for volume change, \textit{i.e}., \textit{C}$_{12}$ (for cubic symmetry). However, the \textit{P$_{C}$} must be positive for the metallic-like bonding, where the electrons are almost delocalized.  For an isotropic crystal, \textit{A${}^{U}$} is zero. The departure of \textit{A${}^{U}$} from zero defines the extent of the elastic anisotropy. In addition to these properties, \textbf{Appendix A} provides a complete list of various parameters and moduli related to the elastic and mechanical properties of materials that  \textsc{\textsc{El\textit{A}Tools}} is able to calculate.

In this work, the relation among the \textit{E}, \textit{G}, \textit{v,} and elastic stiffness constants for a 2D system are derived as,
\begin{equation} \label{42)} 
	\centering
\begin{split}  {E_{x} =\frac{C_{11} C_{22} -C_{12} C_{21} }{C_{22}},} \\ {E_{y} =\frac{C_{11} C_{22} -C_{12} C_{21} }{C_{11} }, }\\ {G_{xy} =C_{66} ,} \\ {\, v_{xy} =\frac{C_{21} }{C_{22},  } , v_{yx} =\frac{C_{12} }{C_{11} } ,}\\  \end{split} 
\end{equation} 
where \textit{E$_{l}$}= \textit{$\sigma$$_{l}$}/\textit{$\varepsilon$$_{l}$} is Young's modulus along the axis of \textit{l}. \textit{v$_{lk}$} =-\textit{d$\varepsilon$$_{k}$}/\textit{d$\varepsilon$$_{l}$} is the Poisson's ratio with tensile strain applied in the \textit{l} direction and the response strain in the \textit{k} direction. \textit{G$_{xy}$} is the shear modulus in the \textit{xy}-plane.

\section{Software description and Features} \label{section:3}
\subsection{Workflow and structure of \textsc{El\textit{A}Tools}}
\begin{figure}
	\includegraphics[scale=0.9]{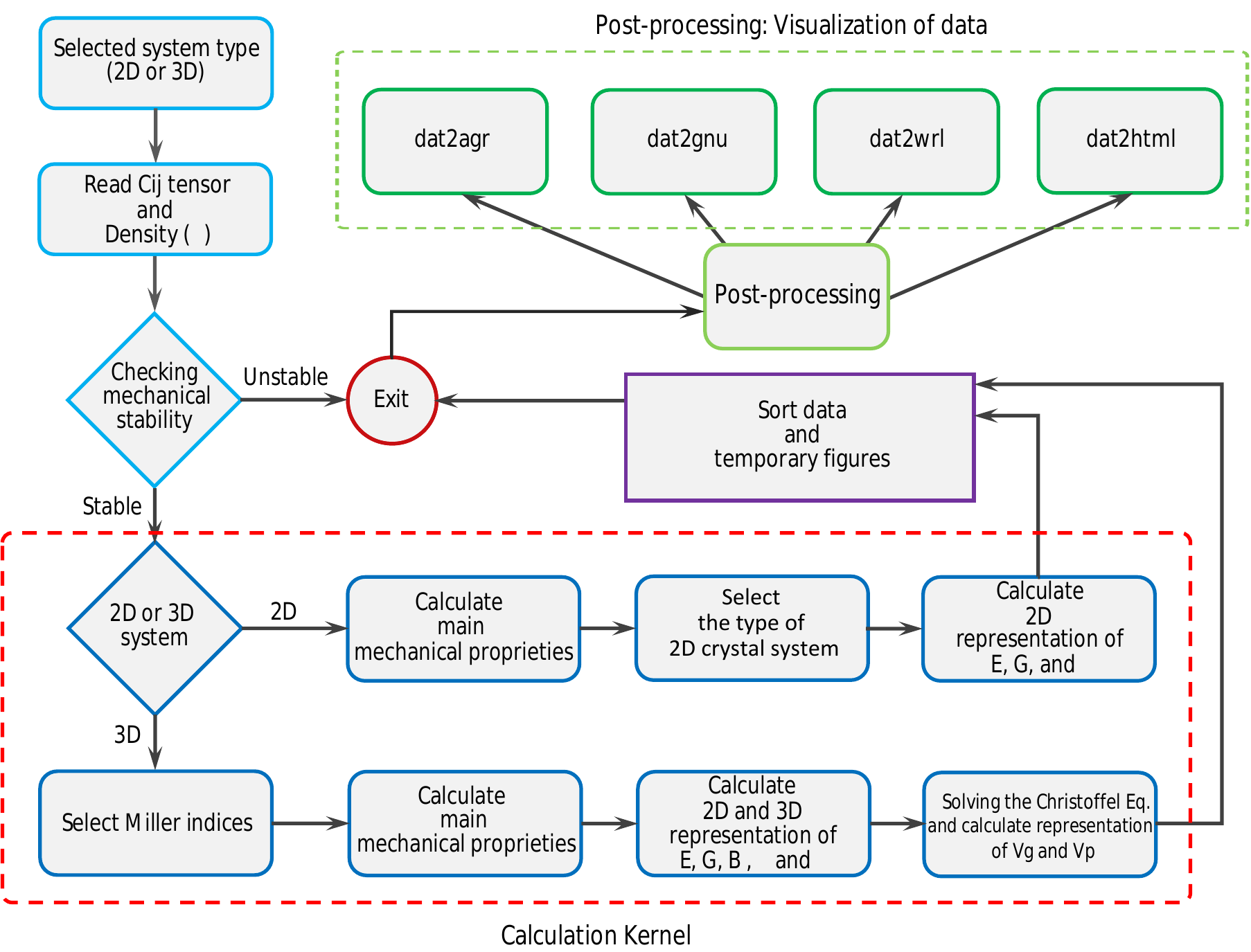} 
	\caption{\label{fig:wide_3}The flowchart of \textsc{El\textit{A}Tools}. The red dashed line represents the Calculation Kernel (CK) block. The green dashed line represents the post-processing stage.}
\end{figure}
The simple workflow of the \textsc{El\textit{A}Tools} package is illustrated in Fig.~\ref{fig:wide_3}. First, the type of either 2D or 3D material is chosen to determine the calculations in the Calculation Kernel (CK) block. Then, \textsc{El\textit{A}Tools} reads the \textit{C$_{ij}$} data as an input file. At this point, \textit{C$_{ij}$} data can be extracted from the output files of IRelast \cite{R18}, IRelast2D \cite{R64}, Elast \cite{R17}, AELAS \cite{R44}, and ElaStic \cite{R19} packages. The output files of these packages are supported as input files by \textsc{El\textit{A}Tools} (see Sec. 3.3). As mentioned earlier, for 3D materials, \textsc{El\textit{A}Tools} has an offline/online database of more than 13000 elastic tensors taken from Materials API (Materials Project database). The user can enter the next steps of calculating \textsc{El\textit{A}Tools} by entering the Materials API-ID of the structure. The developers of \textsc{ElATools} intend to keep the elastic tensors database updated according to the latest release of the Materials Project database. Subsequently, the \textit{C$_{ij}$} tensor enters the mechanical stability check stage. There are four methods of the Born elastic stability conditions for a crystal, which are valid regardless of the crystal symmetry: 1) if the second-order elastic \textit{stiffness} tensor \textit{C$_{ij}$} is definitely positive, 2) if all eigenvalues of \textit{C$_{ij}$} are positive, 3) if all the leading principal minors of \textit{C$_{ij}$} are positive, and 4) if an arbitrary set of minors of \textit{C$_{ij}$} are all positive. Method (2) is used in the \textsc{El\textit{A}Tools}. If the stability conditions are satisfied, the tensor will enter the CK block; otherwise, the program will stop by showing a mechanical instability error.

In the CK block, the calculations are divided into two branches. If the system is 2D the main mechanical properties such as \textit{E$_{x}$}, \textit{E$_{y}$}, \textit{G}$_{xy}$, \textit{v$_{xv}$}, \textit{v}$_{yx}$, etc. are determined, and then enter the orientation-dependent (OD) calculation step. At this stage, according to the equations of Sec. 2.2, OD of Young's modulus, shear modulus, and Poisson's ratio in the (001) plane are calculated. Also, in the (001) plane, the maximum and minimum values for these proprieties are calculated. For the 3D system, the process is like a 2D system. First, \textsc{El\textit{A}Tools} reads the \textit{hkl}-index plane (\textit{e.g.} (100)) entered by the user. Then, the polycrystalline Young's modulus, bulk modulus, shear modulus, P-wave modulus, Poisson's ratio, and Pugh ratio are calculated using three averaging approaches of Voigt, Reuss and Hill approximations. Besides anisotropy indices, Cauchy pressures and hardness information are determined. Subsequently, we arrive at the spatial dependence (SD) calculation process. The SD and the 2D projection of Young's modulus, bulk modulus, shear modulus, Poisson's ratio, and linear compressibility are calculated. To calculate the elastic waves properties such as phase and group velocities, \textsc{El\textit{A}Tools} can solve the \textit{Christoffel} equation at the user's request. Then, similar to the previous steps, it enters the spatial calculation process, and SD and the 2D projection of the phase and group velocities (primary and two secondary modes) are calculated. 

Finally, the calculations obtained from these two branches are sorted and saved. The main mechanical properties are sorted in \textsf{DATA.out}. \textsc{El\textit{A}Tools} creates two directories named \textsf{DatFile-hkl} and \textsf{PicFile-hkl}, and stores files in ``dat'' format, and temporary figures of the properties in these directories, respectively. Temporary figures help us get an overview of the properties. Then, we enter the post-processing for visualization of the 3D spherical plot and their 2D projections with higher quality and detail.

There are four plugins for visualizing data in the post-processing stage: \textsf{dat2gnu.x}, \textsf{dat2agr.x}, \textsf{dat2wrl.x}, \textsf{dat2html.x}. The \textsf{dat2gnu.x} and \textsf{dat2agr.x} generate files for 2D graphical representations of elastic properties in ``gpi'' and ``agr'' formats, respectively, which can be run by G{\scriptsize NUPLOT} and X{\scriptsize MGRACE} programs. The \textsf{dat2wrl.x} and \textsf{dat2html.x} are also prepared for 3D graphical representations of elastic properties, capable of producing files in ``wrl" and ``html" formats. The wrl format can be visualized and explored with a VRML capable browser, such as View3dscene \cite{R35}. For HTML format can also be used from any Web browser. JavaScript written in this format uses \textsf{plotly.js} \cite{R65}, a free and recently open-sourced graphing library. We can represent dynamic parametric surfaces with these formats, making the spatial representation of mechanical properties more straightforward and fully interactive.

 \subsection{Installation and Requirements}
The \textsc{El\textit{A}Tools} is written in \textsf{Fortran90} and is installed with Intel Fortran (ifort) or GNU Fortran (gfortran) compiler. Before installing \textsc{El\textit{A}Tools}, the following libraries and packages should be installed: G{\footnotesize NUPLOT}, and LAPACK (Linear Algebra Package). LAPACK libraries are for numerical calculations and are used to calculate elastic compliance constant and so on. G{\scriptsize NUPLOT} is also used to plot temporary figures and the post-processing stages. One of the packages IRelast, Elast, AELAS, and ElaStic can calculate the elastic \textit{stiffness} constant (\textit{C$_{ij}$}). \textsc{El\textit{A}Tools} supports the output of these packages.

The \textsc{El\textit{A}Tools} is distributed in a compressed tar file \textsf{elatools\_1.**.tar.gz}, which uncompresses into several directories: \textsf{soc}, \textsf{doc}, \textsf{db}, and \textsf{bin}. The \textsf{soc} directory contains the \textsf{f90} files and \textsf{Makefile}. For the compilation, the \textsf{Makefile} must be modified for one's system. The \textsf{doc} directory contains a copy of the short user guide and the examples directory. The \textsf{db} directory contains the elastic constant database files. The path of these files must be specified before installation. More details are provided in the short user guide. After installation, the executable files (\textsf{Elatools.x}, \textsf{dat2gnu.x}, \textsf{dat2agr.x}, \textsf{dat2html.x},  and \textsf{dat2wrl.x}) are saved in the \textsf{bin} directory. Finally, the code will run by executing \textsf{Elatools.x}. 
\subsection{Input and Output files}
The only input data for \textsc{El\textit{A}Tools} is the elastic \textit{stiffness} constant, which can be calculated by other packages. However, \textsc{El\textit{A}Tools} also supports output files of many packages for convenience, such as IRelast (or IRelast2D) with \textsf{INVELC-matrix} output file, Elast with \textsf{elast.output} output file, AELAS with \textsf{ELADAT} output file, ElaStic with \textsf{ElaStic\_2nd.out} output file, and \textsf{Cij.dat} (3D system) or \textsf{Cij-2D.dat} (2D system) file for any other outputs (See \textbf{Appendix B} for more details). Several output files are generated in each run:
\begin{itemize}
	\item  Spatial-dependence and 2D projection files in 3D materials.  
	
	For this case, fifteen files \textsf{3d\_pro.dat} (\textsf{pro=bulk}, \textsf{young}, \textsf{poisson}, \textsf{comp}, \textsf{shear}, \textsf{pp}, \textsf{pf}, \textsf{ps}, \textsf{gp}, \textsf{gf}, \textsf{gs}, \textsf{pfp}, \textsf{pff}, \textsf{pfs}, \textsf{km}) are generated. Among these files, the \textsf{3d\_poisson.dat} file includes the maximum value, minimum positive value, minimum negative value, and average value of Poisson's ratio. The \textsf{3d\_shear.dat} file contains the maximum positive value, minimum positive value, and average value of shear modulus, and the \textsf{3d\_comp.dat} file also contains the positive and negative value of linear compressibility. Also, for 2D projection of any plane, nine files \textsf{2dcut\_pro.dat} (\textsf{pro=bulk, young, poisson, comp }, \textsf{shear}, \textsf{km}, \textsf{gveloc}, \textsf{pveloc}, and \textsf{pfaveloc}) are generated. It should be noted that \textsf{2dcut\_p/g/pfveloc.dat} files include primary, fast secondary, and slow secondary modes. See Table \ref{Tab:2} for more details on \textsf{3d\_pro.dat} and \textsf{2dcut\_pro.dat} files. 
	
	\item  Orientation-dependent files in 2D materials.
	
	In this case, three files \textsf{pro} (\textsf{pro= young, poisson, shear}) are generated. Also, the \textsf{poisson\_2d\_sys.dat} file contains the maximum value, the minimum positive value, and the minimum negative value of Poisson's ratio.
	
	\item  \textsf{DATA.dat} file.
	
	This file contains the \textit{C$_{ij}$}, \textit{S$_{ij}$}, the main properties and the minimal and maximal values of Young's modulus, bulk modulus, shear modulus, Poisson's ratio, linear compressibility, power flow angle, phase, and group velocities as well as, the angles and directions along which these extrema occur.
	
	\item  Temporary files.
	
	These files are used for post-processing.
\end{itemize}
\subsection{ Visualization and Post-processing}
\begin{figure}
	\includegraphics[scale=1]{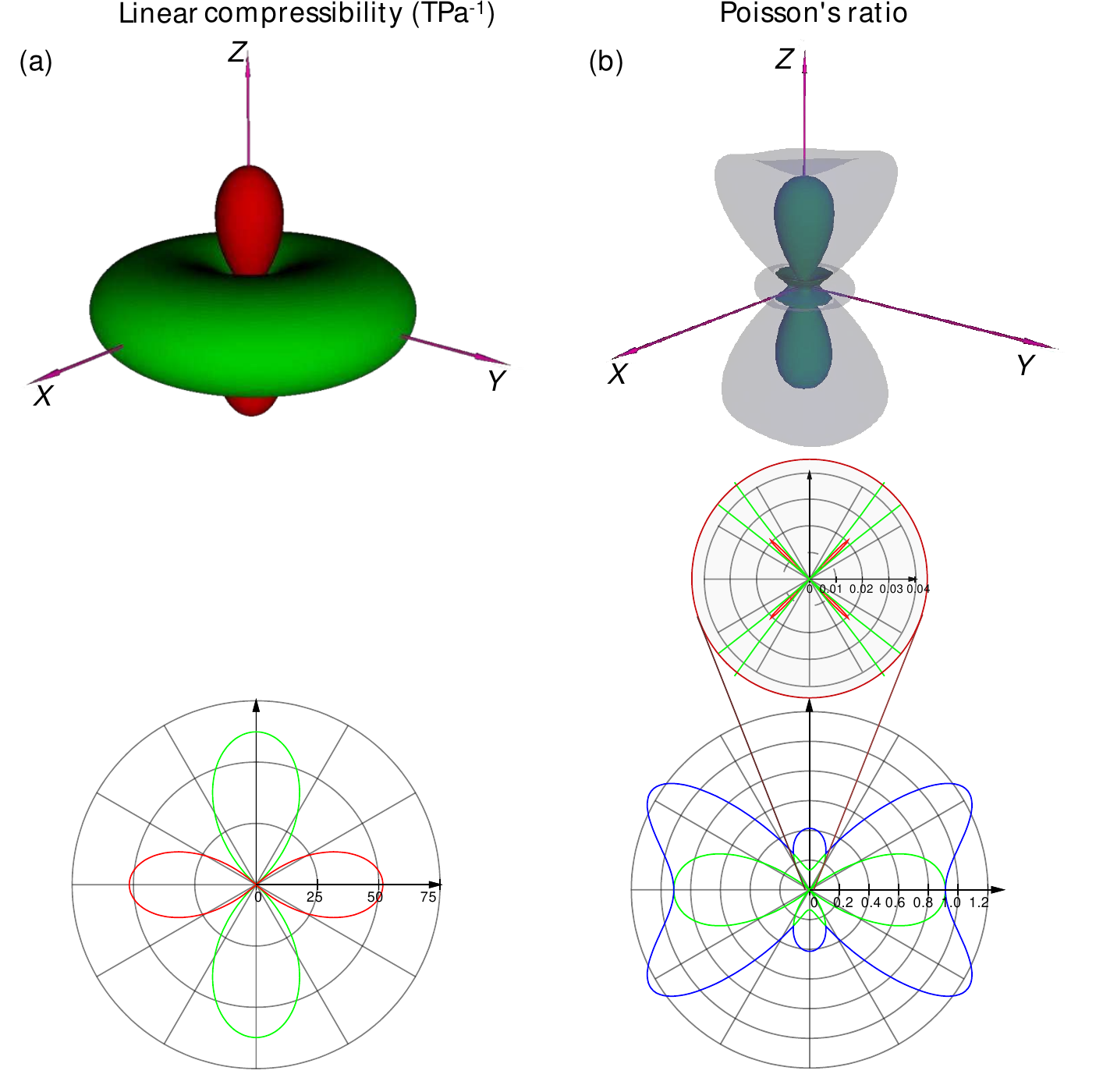} 
	\caption{\label{fig:wide_4}(a) The spatial-dependence and (b) 2D projection in (110) plane of linear compressibility and Poisson's ratio of ZnAu$_{2}$(CN)$_{4}$ structure.}
\end{figure}

\begin{figure}
	\includegraphics[scale=0.2]{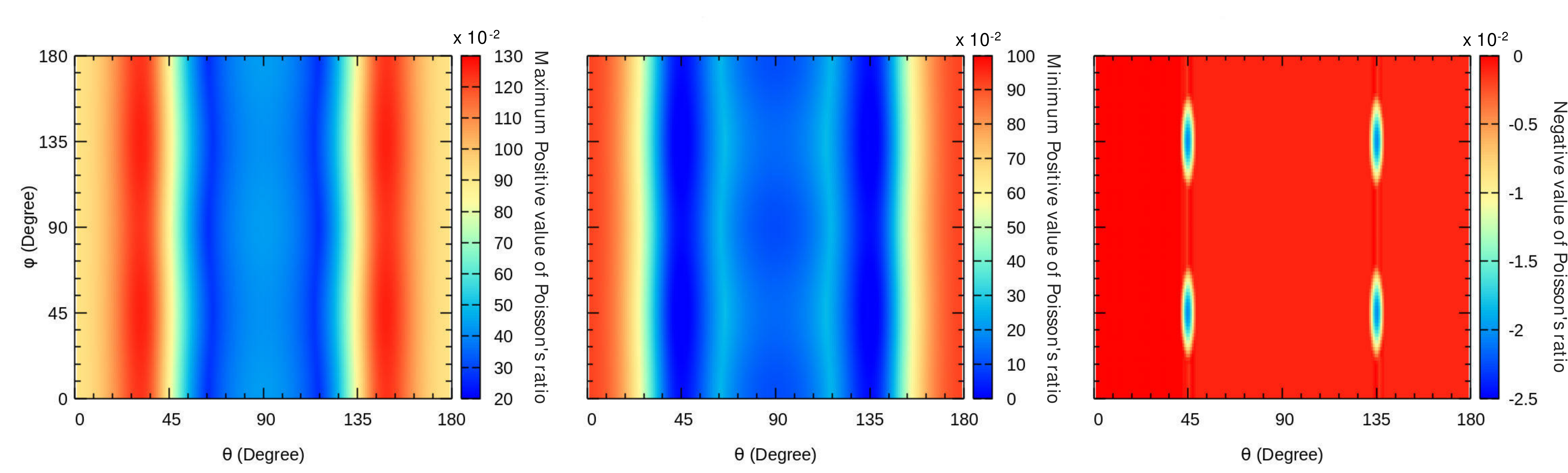}
	\caption{\label{fig:wide_4-2}Poisson's ratio heat maps with respect to $\theta$ and $\phi$ angles for ZnAu$_{2}$(CN)$_{4}$ compound. (a) Maximum  positive values, (b) minimum positive values, and (c) negative values of Poisson's ratio.}
\end{figure}
\begin{figure}
	\includegraphics[scale=0.8]{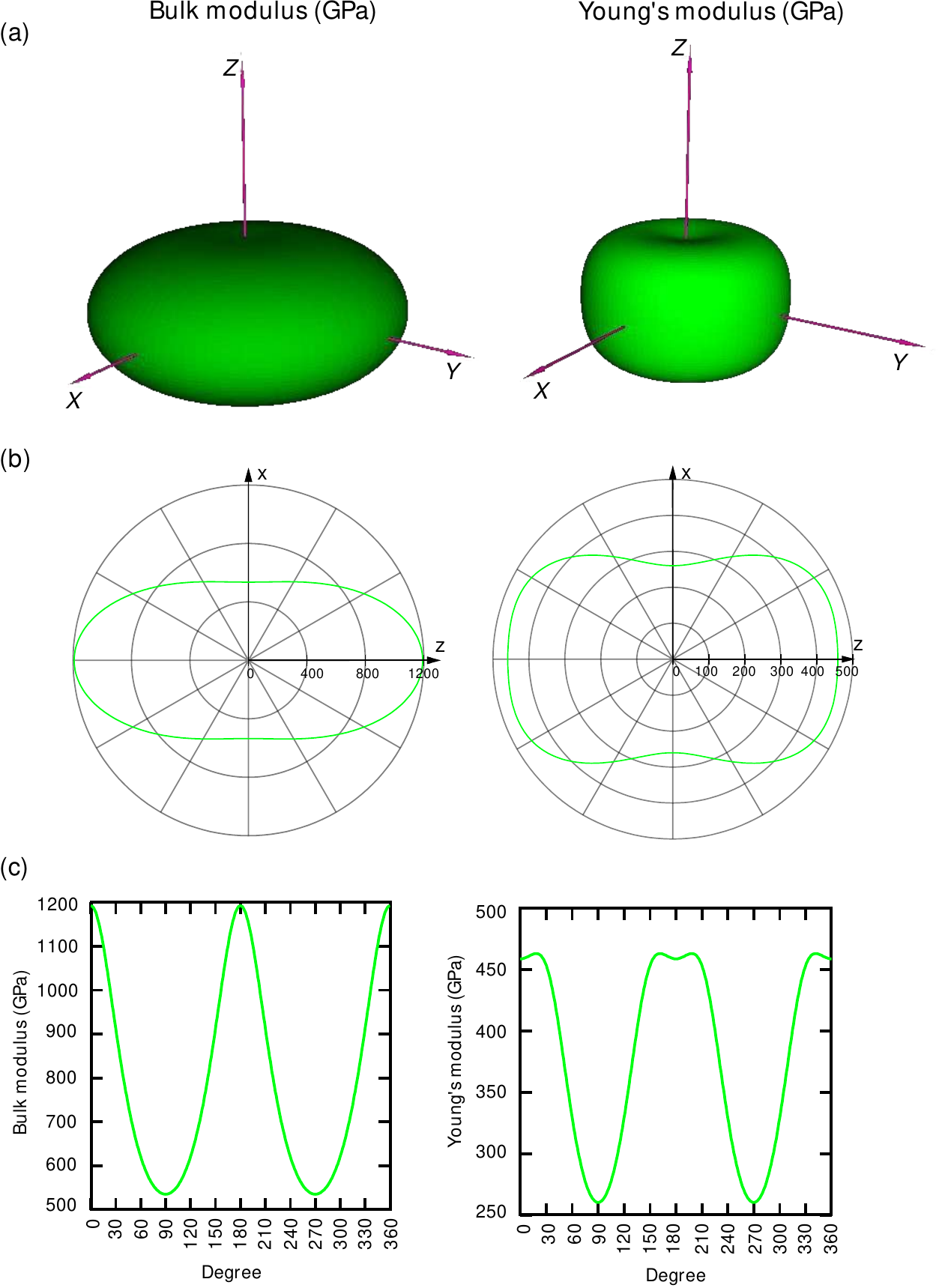} 
	\caption{\label{fig:wide_5} (a) The spatial-dependent and 2D projection in (b) polar and (c) cartesian coordinates in (110) plane of Poisson's ratio and linear compressibility of the CrB$_{2}$ structure.}
\end{figure}
\begin{figure}
	\includegraphics[scale=0.9]{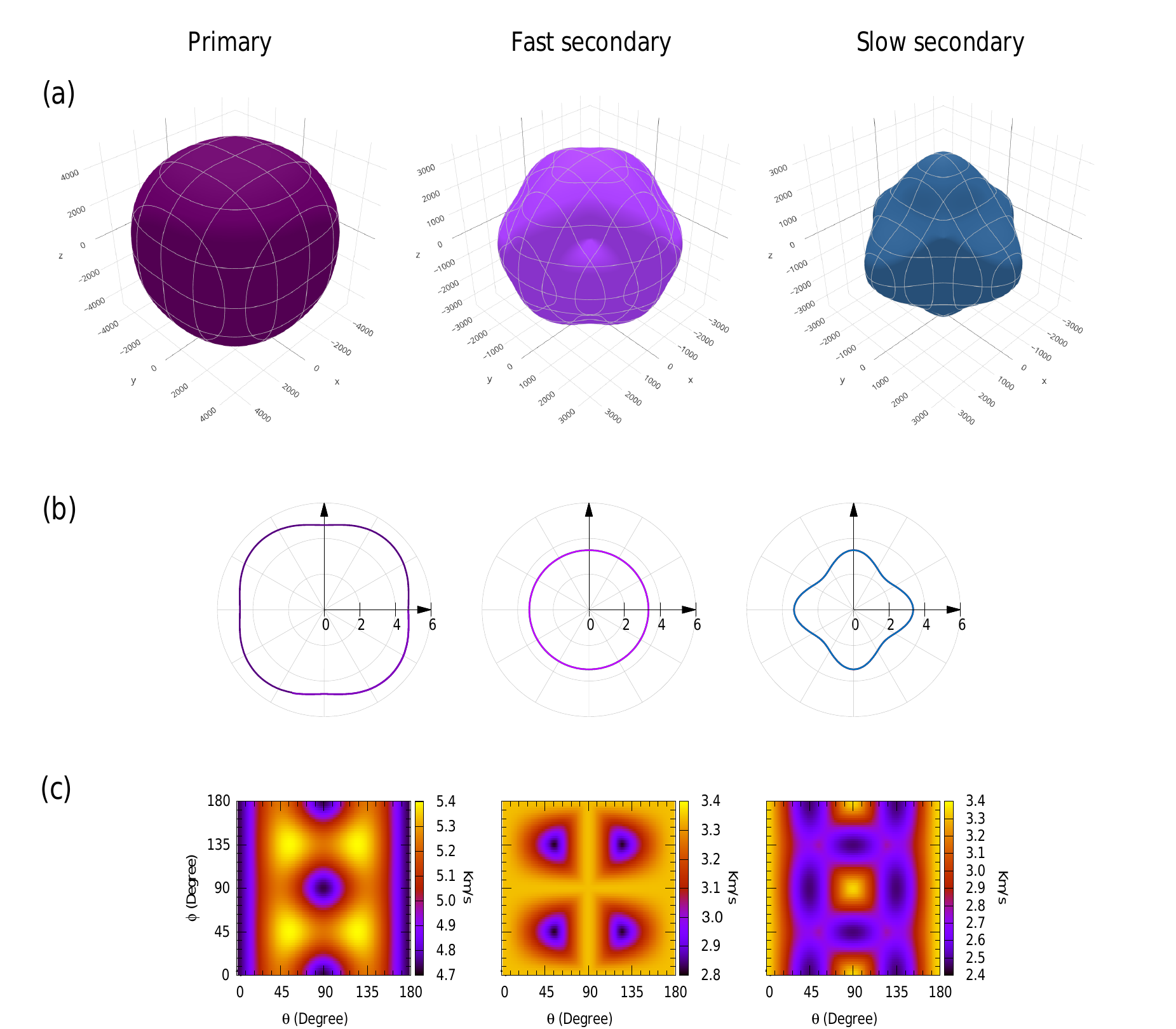} 
	\caption{\label{fig:wide_61}(a) The spatial-dependent, (b) 2D projection in polar coordinates on (100) plane, and (c) heat maps of phase velocity ($\nu_{p}$) for primary, fast, and slow secondary modes of the GaAs compound.}
\end{figure}
\begin{figure}
	\centering
	\includegraphics[scale=0.9]{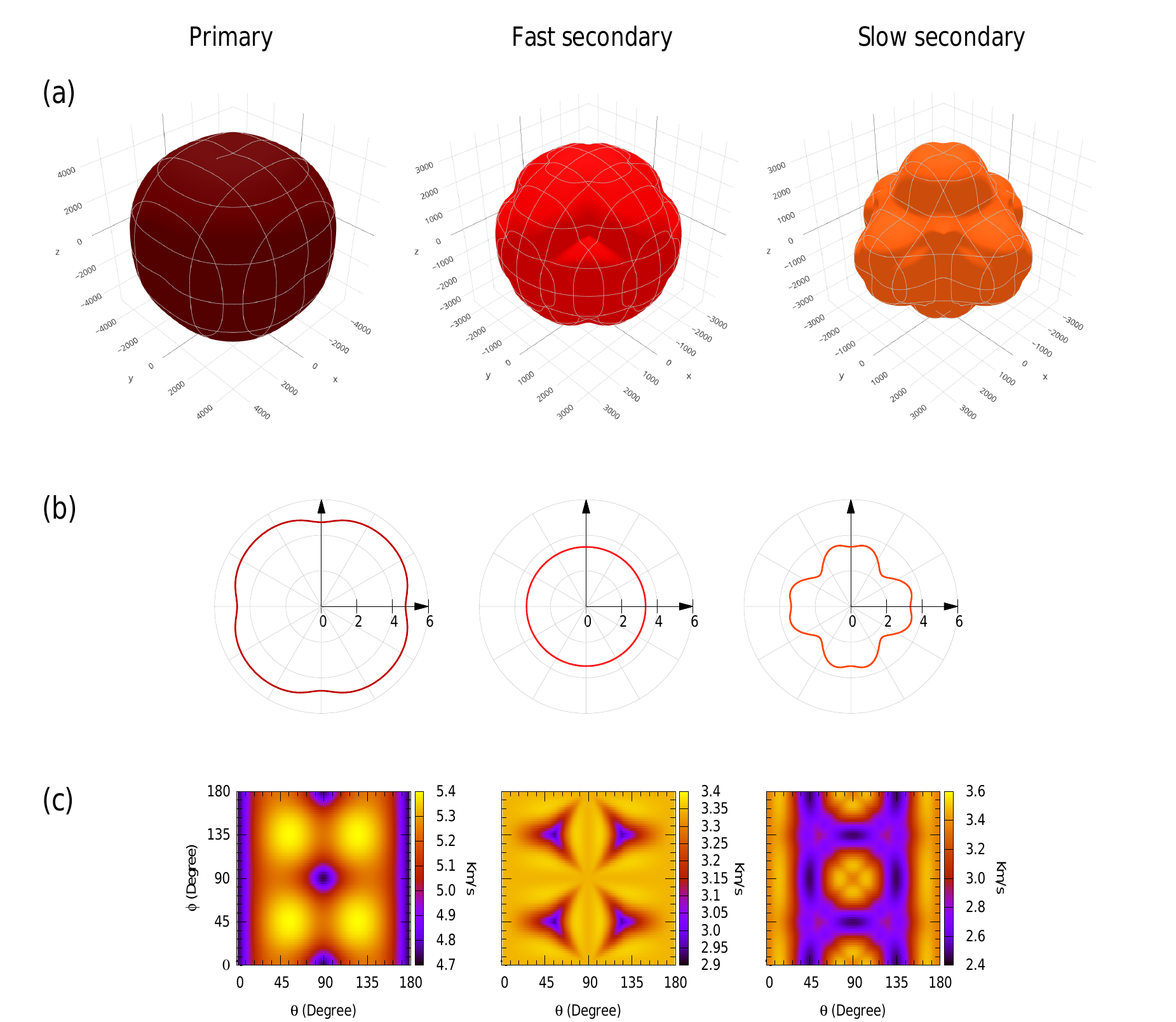} 
	 \caption{\label{fig:wide_62}(a) The spatial-dependent, (b) 2D projection in polar coordinates on (100) plane, and (c) heat maps of group velocity ($\nu_{g}$) for primary, fast, and slow secondary modes of the GaAs compound.}
\end{figure}
\begin{figure}
	\includegraphics[scale=0.9]{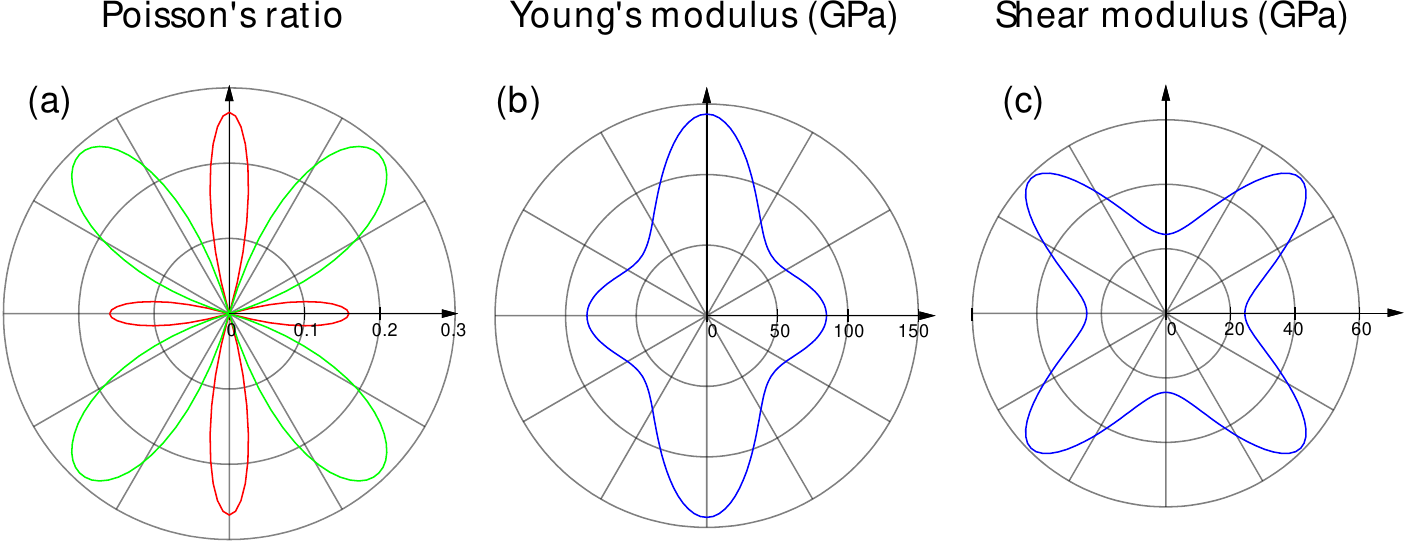} 
	\caption{\label{fig:wide_6}The orientation-dependent in-plane (a) of Poisson's ratio, (b) Young's modulus, and (c) shear modulus of \textit{$\delta$}-phosphorene.}
\end{figure}
\begin{figure}
	\includegraphics[scale=1.2]{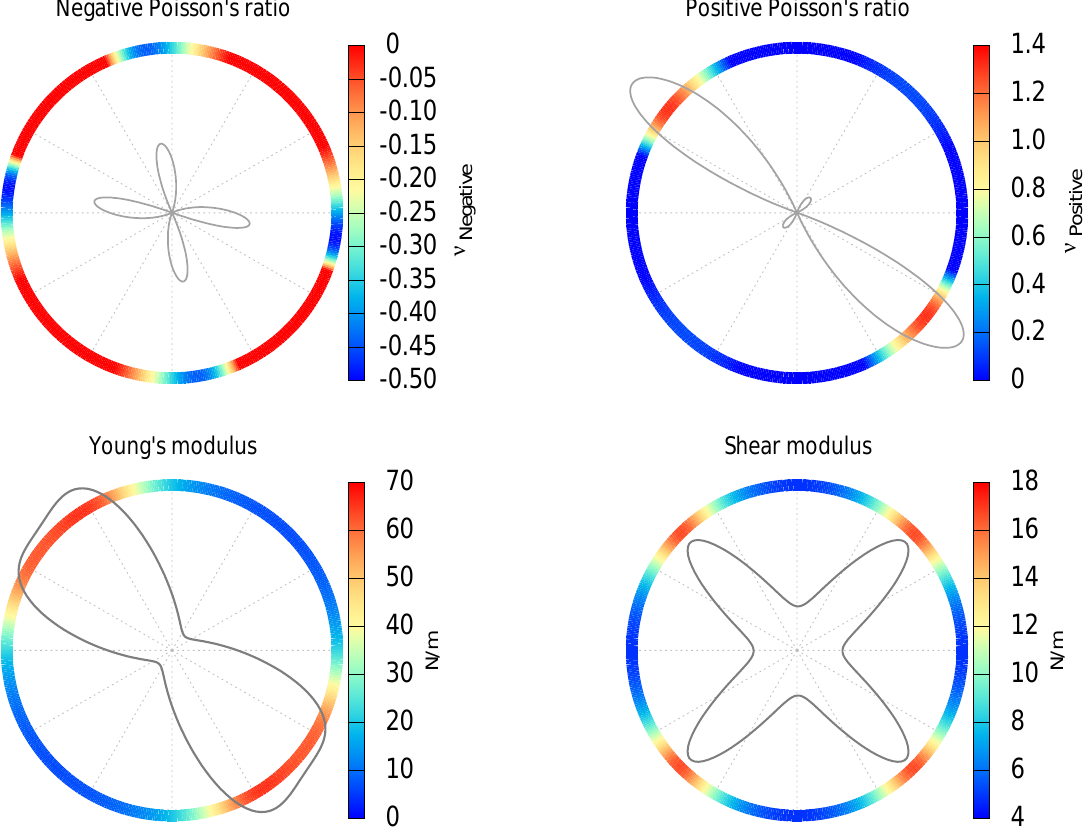} 
	\caption{\label{fig:wide_10} The polar heat maps of Poisson's ratio, Young's modulus, and shear modulus of Pd$_{2}$O$_{6}$Se$_{2}$ monolayer}
\end{figure}
\begin{figure}
	\includegraphics[scale=0.8]{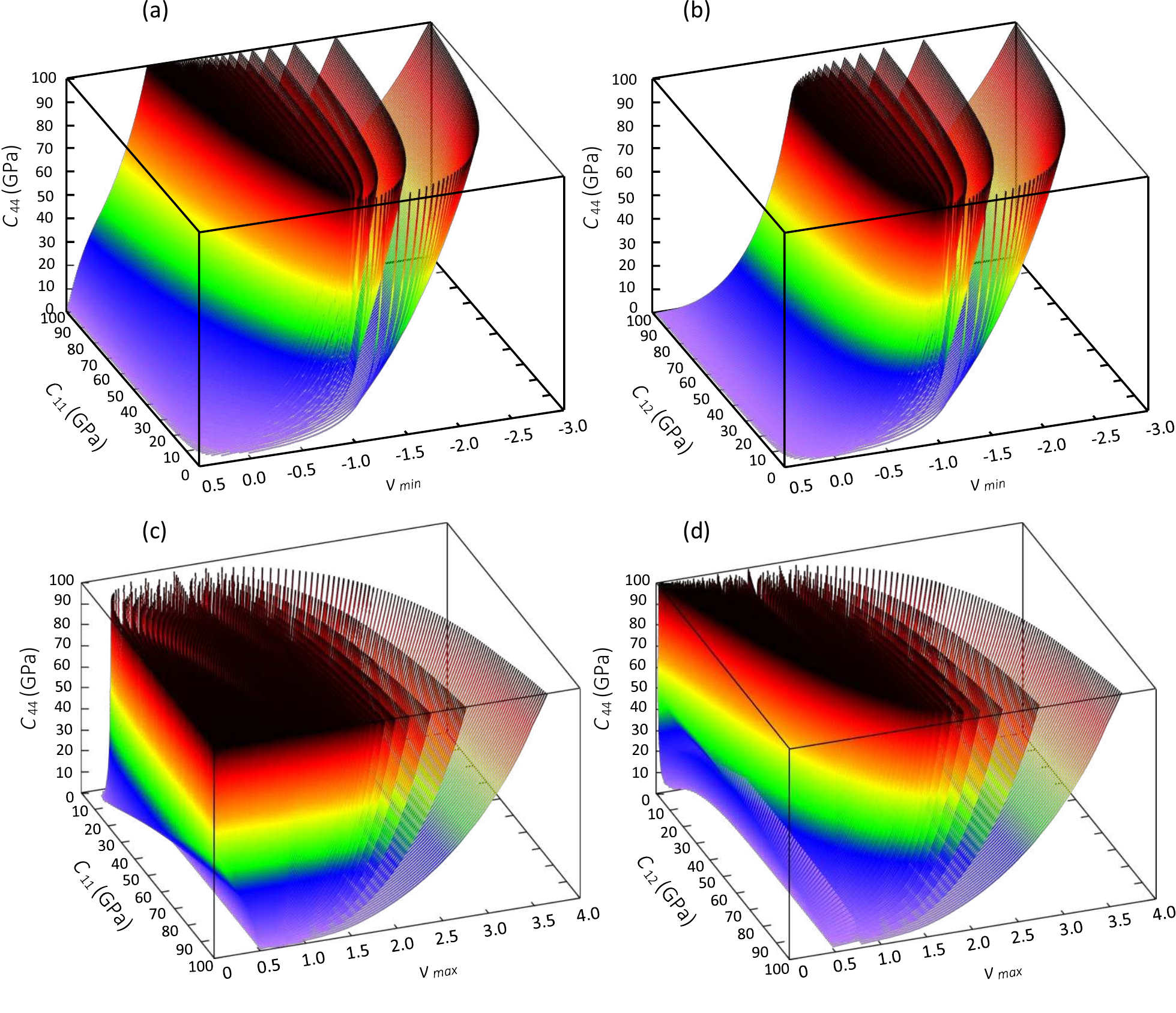} 
	\caption{\label{fig:wide_7}The minimum and maximum values of Poisson's ratio diagram with respect to (a, b) (\textit{C}$_{44}$, \textit{C}$_{12}$) and (c, d) (\textit{C}$_{44}$, \textit{C}$_{11}$).}
\end{figure}
\begin{figure}
	\includegraphics[scale=0.8]{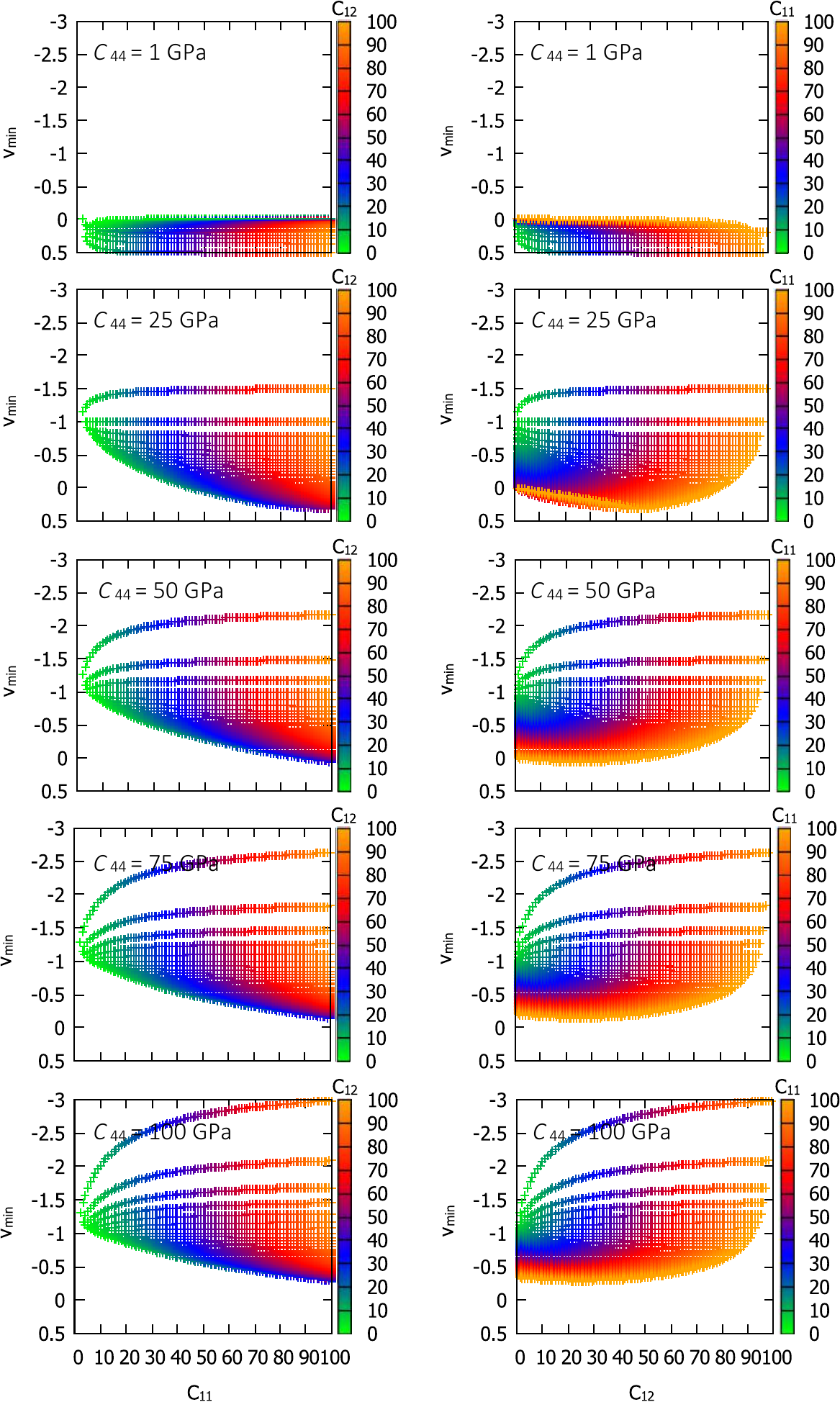} 
	\centering
	\caption{\label{fig:wide_8}Five slices at constant values of \textit{C}$_{44}$ in which the \textit{v$_{min}$} is a function of \textit{C}$_{11}$ and \textit{C}$_{12}$ coefficients.}
\end{figure}
\begin{figure}[H]
	\includegraphics[scale=0.8]{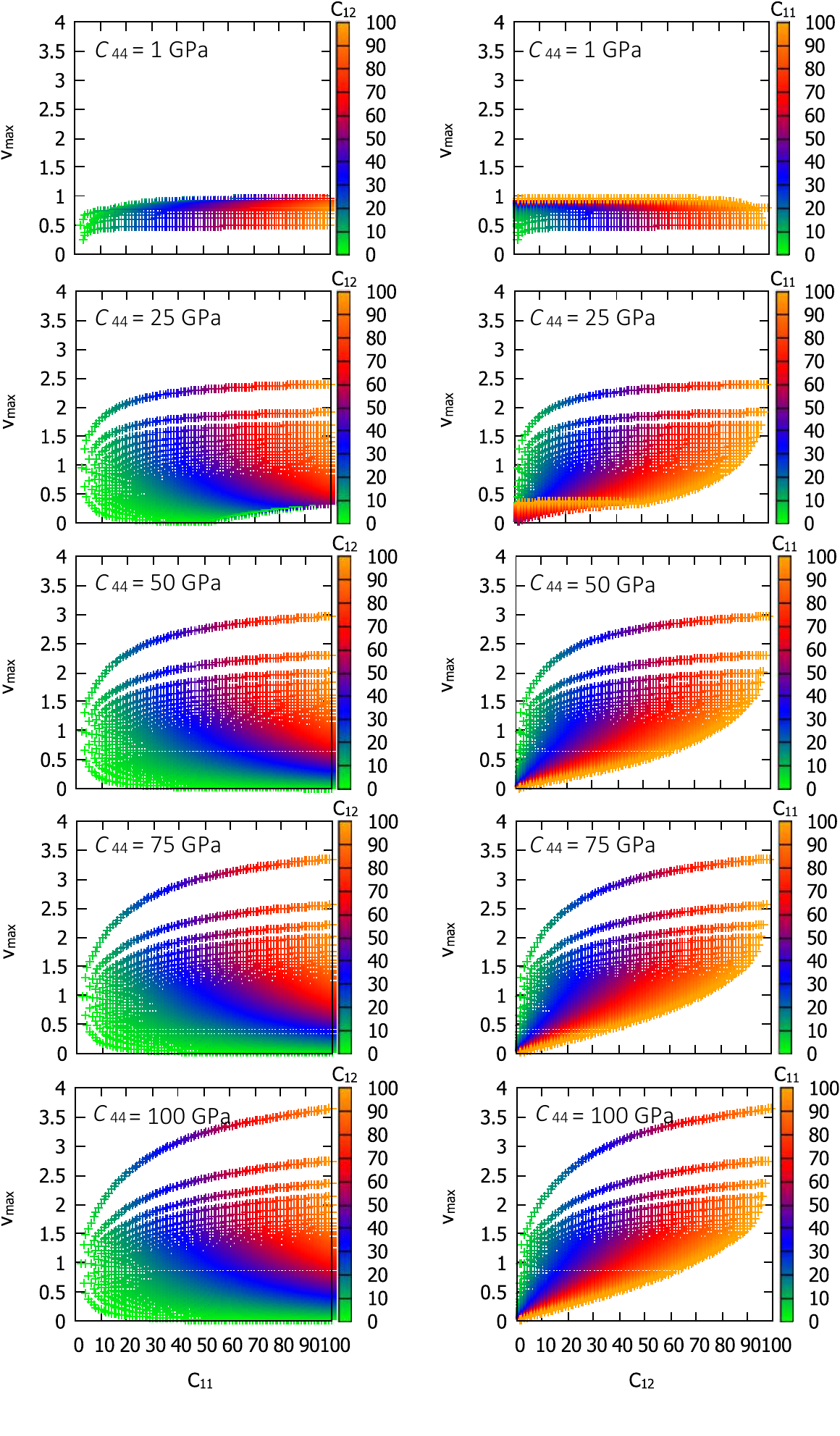} 
	\centering
	\caption{\label{fig:wide_9}Five slices at constant values of \textit{C}$_{44}$ in which the \textit{v$_{max}$} is a function of \textit{C}$_{11}$ and \textit{C}$_{12}$ coefficients.}
\end{figure}

In the post-processing, three powerful tools \textsf{dat2gnu.x}, \textsf{dat2agr.x}, \textsf{dat2wrl.x}, and \textsf{dat2html.x} are designed to visualize of results, with output files in gpi, agr, wrl, and html formats, respectively. In Figs.~\ref{fig:wide_4}--\ref{fig:wide_9}, we show the corresponding plots for ZnAu$_{2}$(CN)$_{4}$ (space group P6222) \cite{R45}, CrB$_{2}$ (space group P6/mmm) \cite{R46}, GaAs (space group F-43m) \cite{R85}, (space group C2/m) \cite{R84}, {$\delta$}-phosphorene (space group Pmc21), and Pd$_{2}$O$_{6}$Se$_{2}$ (monolayer) structures \cite{R47} by these three postprocessing tools and G{\footnotesize NUPLOT}, X{\scriptsize MGRACE} and view3dscene programs. In the following section, these three structures are examined. A list of the main elastic properties and anisotropy indices of ZnAu$_{2}$(CN)$_{4}$, CrB$_{2}$, GaAs,  $\delta$-phosphorene ($\delta$-P), and Pd$_{2}$O$_{6}$Se$_{2}$ monolayer are given in \textbf{Appendix C}.

\textbf{Test case (1): ZnAu$_{2}$(CN)$_{4}$}. Fig.~\ref{fig:wide_4} shows the spatial-dependence and 2D projection in (110) plane of Poisson's ratio and linear compressibility of ZnAu$_{2}$(CN)$_{4}$ structure. The negative linear compressibility for this compound is shown in Fig.~\ref{fig:wide_4}(a). In these Figs., directions corresponding to positive values of linear compressibility are plotted in green, and those of NLC are plotted in red. The NCL of ZnAu$_{2}$(CN)$_{4}$  was predicted in Ref. \cite{R45}, which is evident in the \textit{z}-direction. The spatial-dependence and 2D projection in (110) plane of Poisson's ratio are shown in Fig.~\ref{fig:wide_4}(b). For Poisson’s ratio, which can be negative in some directions, three categories of colors are considered: directions corresponding to maximum (minimum) positive values of Poisson’s ratio are plotted in translucent blue (green) color, and those of NPR are plotted in red color. Calculations of \textsc{El\textit{A}Tools} based on the elastic tensor in Ref. \cite{R45} show that this structure, in addition to the NLC has a small NPR (-0.02) on (110) plane. Using G{\footnotesize NUPLOT} and \textsf{dat2gnu.x} tools,  heat maps with respect to $\theta$ and $\phi$ angles are shown in Fig.~\ref{fig:wide_4-2}.  These 2D heat maps show the changes in the Poisson's ratio relative to $\theta$ and $\phi$ angles in spherical space. NPR is well visible in the Fig.~\ref{fig:wide_4-2}(c). For comparison, ELATE display the NPR feature due to the unavailability 2D representation in (110) plane (it can display only on three planes (100), (010), and (001)). Hence, the ability to select a custom plane is a
unique feature in \textsc{El\textit{A}Tools}. 

\textbf{Test case (2): CrB$_{2}$}.\textbf{ }CrB$_{2}$ compound is investigated to evaluate \textsc{El\textit{A}Tools} and post-processing. Elastic tensor is taken from the calculations of Ref. \cite{R46}. The Young's modulus and bulk modulus are shown in 3D and 2D ((010) plane) in Fig.~\ref{fig:wide_5}. The spatial dependent files are generated by the \textsf{dat2wrl.x} and represented by the View3dscene programs (Fig. ~\ref{fig:wide_5}(a)). The orientation-dependent files in polar (Cartesian) coordinate are generated by \textsf{dat2gnu.x} (\textsf{dat2agr.x}) and displayed by G{\footnotesize NUPLOT} (X{\scriptsize MGRACE}) (see Fig.\ref{fig:wide_5}(b) and (c)). For an isotropic system, in the spatial-dependence (polar and cartesian coordinates), the graph would be a sphere (a circle and a straight line). Fig.~\ref{fig:wide_5}(a) shows that the bulk modulus and Young’s modulus of CrB$_{2}$ have anisotropy. The projections on the (010) plane show more details about the anisotropic properties of the bulk modulus and Young's modulus.

\textbf{Test case (3): GaAs.} We employ gallium arsenide (GaAs) as an example to illustrate the capabilities of \textsc{ElaTools}, and the output figures will be briefly commented on here. In this example, we investigate the elastic wave properties of this compound. The values of $C_{ij}$ and $\rho$ were taken from Ref. \cite{R85}. In Figs. \ref{fig:wide_7} and \ref{fig:wide_8}, phase and group velocities for primary, fast, and slow secondary modes are calculated by \textsc{Elatools} and obtained by the \textsf{dat2html.x} (using plotly.js), \textsf{dat2gnu.x} (using \textsc{Gnuplot}) post-processing codes. In Figs. \ref{fig:wide_61} and \ref{fig:wide_62} it is clear that the distinction between fast secondary (FS) and slow secondary (SS) modes always refers to the phase velocity since the group velocity of the FS mode could be lower than that of the SS mode for certain propagation directions. According to Table \ref{Tab:8} and these Figures, the minimum and maximum anisotropy are associated with primary (P) and SS modes, respectively. In Figs. \ref{fig:wide_61}(a) and \ref{fig:wide_62}(a), the P modes are not spherical, and their anisotropy is higher than the other modes. Also, as shown in Figs. \ref{fig:wide_61}(c) and \ref{fig:wide_62}(c), the P modes propagation patterns are more complex, indicating higher anisotropy than other modes.

\textbf{Test case (4): \textit{$\boldsymbol{\delta}$}-Phosphorene}.  Haidi Wang \textit{et al.} \cite{R47} have discovered that \textit{$\delta$}-phosphorene is a superior 2D\textit{ auxetic} material with high NPR. In Fig.~\ref{fig:wide_6}, the Poisson's ratio, shear modulus, and Young's modulus are calculated by \textsc{El\textit{A}Tools} and shown by the \textsf{dat2gnu.x} and G{\footnotesize NUPLOT}. As shown in Fig.~\ref{fig:wide_6}(a), the maximum value of the NPR (-0.267) occurs at a 90-degree angle. This amount is in perfect agreement with Haidi Wang \textit{et al.} Also, the orientation-dependent Young's modulus of this structure (see Fig.~\ref{fig:wide_6}(b)) is in good agreement with Haidi Wang \textit{et al.} \textsc{El\textit{A}Tools} can calculate shear modulus in 2D materials. Fig.\ref{fig:wide_6}(c) shows this feature of \textit{$\delta$}-phosphorene.

\textbf{Test case (5): Pd$_{2}$O$_{6}$Se$_{2}$ monolayer}. In this test case, the mechanical properties of the Pd$_{2}$O$_{6}$Se$_{2}$ monolayer with an oblique 2D crystal system are investigated. The values for the $C_{ij}$ was taken from computational 2D materials database (C2DB) \cite{R47}. In Fig.~\ref{fig:wide_10}, the polar heat maps of Poisson's ratio, Young's modulus, and shear modulus are calculated by \textsc{El\textit{A}Tools} and shown by the \textsf{dat2gnu.x} and \textsc{Gnuplot}. The main elastic properties and anisotropy indices of this monolayer are listed in Table \ref{Tab:9}. It is clear from this table that the Pd$_{2}$O$_{6}$Se$_{2}$ monolayer has a NPR (-0.49). This result is well recognizable by polar heat maps (see Fig. \ref{fig:wide_10}). Polar heat maps in visualizing the mechanical properties of 2D materials, such as heat maps in 3D, are useful tools for searching for NPR with small values.

\textbf{Test case (6): NPR analysis of cubic symmetry materials. }Arbitrarily large positive and negative values of Poisson’s ratio could occur in solids with cubic material symmetry \cite{R48,R49}. To investigate this matter, we have calculated the Poisson's ratio of a hypothetical set of systems with cubic symmetry that include three independent elastic coefficients (\textit{C}$_{11}$, \textit{C}$_{12}$, and \textit{C}$_{44}$) in the range between 0 and 100 GPa (with steps of 0.5 GPa), considering the mechanical stability criteria, and using \textsc{El\textit{A}Tools}.

Fig.~\ref{fig:wide_7} shows the minimum (\textit{v$_{min}$}) and maximum (\textit{v$_{m}$$_{ax}$}) values of Poisson's ratio diagram with respect to (\textit{C}$_{44}$, \textit{C}$_{12}$) and (\textit{C}$_{44}$, \textit{C}$_{11}$). As shown in Figs.~\ref{fig:wide_7}(a) and (b), \textit{C}$_{44}$ plays a vital role in the negative values of \textit{v$_{min}$}. Also, comparing these two figures, with increasing \textit{C}$_{44}$, the value of \textit{C}$_{11}$ has a more prominent role than the value of \textit{C}$_{12}$ in NPR. For a better investigation, by combining  Figs.~\ref{fig:wide_7}(a-b) and Figs.~\ref{fig:wide_7}(c-d), 5 slices in constant \textit{C}$_{44}$ are shown in Fig.~\ref{fig:wide_8} and Fig.~\ref{fig:wide_9}. In Fig.~\ref{fig:wide_8}, \textit{v}$_{min}$ values are almost positive when \textit{C}$_{44}$ =1 GPa and \textit{C}$_{11}$ and \textit{C}$_{12}$ range from 1 to 100 GPa. Few materials have been found that have such independent elastic coefficients. When \textit{C}$_{44}$ increases from 1 to 50 GPa, for all \textit{C}$_{11}$ and \textit{C}$_{12}$, \textit{v$_{min}$} changes its sign from positive to negative. As can be seen, when \textit{C}$_{44}$ reaches 100 GPa, the maximum negative value of \textit{v$_{min}$} is -3. In Fig.~\ref{fig:wide_9}, when \textit{C}$_{44}$ = 1 the maximum value of \textit{v$_{max}$} is less than one, and with increasing \textit{C}$_{44}$, the \textit{v$_{max}$} increases and can reach 4. As shown in both Figs. \ref{fig:wide_8} and~\ref{fig:wide_9}, the patterns of changes in \textit{v$_{min}$} and \textit{v$_{max}$} are the same when \textit{v$_{min}$} $\mathrm{<}$ -1 and \textit{v$_{max}$}$\mathrm{>}$\textit{ }1. In general, it can be concluded that the coefficient of \textit{C}$_{44}$ has a more critical role than the other two coefficients (\textit{C}$_{11}$ and \textit{C}$_{12}$) in the NPR of materials. These negative values of \textit{v$_{min}$} also appear when stretched along the [110] direction.There are many compounds of cubic symmetry that can be placed in this range of elastic coefficients that can have NPR.

Finally, we have prepared a documentation \href{https://yalameha.gitlab.io/elastictools/}{website} that provides more examples and tutorials for \textsc{El\textit{A}Tools}.

\section{Summary and outlook}\label{section:4}
We introduced \textsc{El\textit{A}Tools}, a \textsf{Fortran90} code designed to analyze the second-order elastic tensors of three and two-dimensional crystal systems. \textsc{El\textit{A}Tools} offers a helpful tool for detecting elastic anisotropy, NLC, and NPR or \textit{auxetic} materials. Four post-processing programs specifically designed for the visualization of the results are provided. Besides, \textsc{El\textit{A}Tools} includes the elastic constant database of Materials Project for 3D materials allowing offline/online use. Furthermore, the code can generate data for Machine Learning to detect and predict elastic and anisotropy properties. The authors plan to extend \textsc{El\textit{A}Tools} to analyze other tensorial properties, such as piezoelectric and photoelastic tensors.

  \subsection{Appendix A}
A list of main elastic properties and anisotropy indices of two-dimensional and three-dimensional materials is provided in Table \ref{Tab:1}. The elastic modulus B, E, and G are defined by Eqs.\eqref{35)}, \eqref{36)}, and \eqref{37)}. Isotropic Poisson's ratio and 2D Poisson's ratio in 3D and 2D materials are defined by Eqs.\eqref{37)} and \eqref{42)}, respectively.
  
The P-wave modulus (M), known as the \textit{longitudinal} modulus, is associated with the homogeneous isotropic linear elastic materials. This modulus describes the ratio of axial stress to the axial strain in a uniaxial strain state \cite{R66}, and is defined as follows:
  \begin{equation} \label{43)} 
  	M = B + 4G/3.
  \end{equation}

Pugh's ratio or B/G ratio deﬁnes the ductility or brittleness of a given material. The critical value of Pugh's ratio is found to be 1.75. Materials with B/G $>$ 1.75 are ductile, whereas those with B/G $<$ 1.75 are brittle in nature \cite{R6,R23,R67}.
  
 Lame’s ﬁrst ($\lambda_{1}$) and second ($\lambda_{2}$) parameters help to parameterize Hooke's law in 3D for homogeneous and isotropic materials using the stress and strain tensors. The $\lambda_{1}$ provides a measure of compressibility, and the $\lambda_{2}$ is associated with the shear stiffness of a given material \cite{R66}. These two parameters are specified as follows:
  \begin{equation} \label{44)} 
	\lambda_{1} = \dfrac{\nu E}{(1+\nu)(1-2\nu)},\,\, \lambda_{2}= \dfrac{E}{2(1+\nu)}.
\end{equation} 

Kleinman's parameter ($\xi$) describes the stability of a solid under stretching or bending, and is defined as follows:
  \begin{equation} \label{45)} 
	\xi = \dfrac{C_{11}+8C_{12}}{7C_{11}-2C_{12}}.
\end{equation}
$\xi$ = 1 implies that bond stretching would be dominated, while $\xi$ = 0 implies that bond bending would be dominated.

Thermal conductivity, responsible for conducting heat energy, is a useful physical parameter for practical applications. It decreases with increasing temperature toward a limiting value known as the minimum thermal conductivity ($\kappa_{m}$). The value of $\kappa_{m}$ can be obtained using Cahill \cite{R69} and Clarke \cite{R68} models from the following expressions:
\begin{equation} \label{46)} 
	\kappa_{m}^{Clarke} = 0.87k_{B} M_{a}^{2/3} E^{1/2} \rho^{1/6},
\end{equation}
\begin{equation} \label{47)} 
	\kappa_{m}^{Cahill} = (k_{B}/2.48)n^{2/3} (\nu_{l} + 2\nu_{t} ),
\end{equation}
where $k_{B}$ is Boltzmann's constant, E and $\rho$ are Young's modulus and the density of the material, respectively, and $M_{a}$ is the mean mass of atoms in each unit cell, which can be calculated by $M_{a}$=[M/(m$N_{A}$ )] (M and m are the molar mass and the total number of atoms in each unit cell, respectively, and $N_{A}$ is Avogadro's constant). In Cahill's model, \textit{n} is the density of the atom number per unit volume, and $\nu_{l}$ and $\nu_{t}$ are the longitudinal and transverse sound velocities, respectively (see Eqs. \eqref{63)},\eqref{64)} and \eqref{65)}). Currently, \textsc{ElaTools} calculates the $\kappa_{m}$ value by Clarke's model.

Cauchy's pressure ($P_{C}$) is associated with the angular characteristic of atomic bonding in a given material and is defined in different symmetries by Eqs. \eqref{39)}, \eqref{40)}, and \eqref{41)}.

Elastic anisotropy is an important property to characterize for a comprehensive understanding of the mechanical and physical properties of materials. This property influences a variety of physical processes like geophysical explorations of the Earth's interior \cite{R70}, development of plastic deformation in crystals \cite{R71}, enhanced positively charged defect mobility \cite{R72}, microscale cracking in ceramics \cite{R73}, alignment or misalignment of quantum dots \cite{R74}, etc. Various methods have been reported in the literature to quantify the elastic anisotropy based on elastic modulus and $C_{ij}$ tensor. Ranganathan and Ostoja-Starzewski \cite{R75} derived a universal anisotropy index $A^{U}$ to provide a measure of elastic anisotropy. This index is called \textit{universal} because of its applicability to all crystal symmetries and can be defined as follows \cite{R75}:
\begin{equation} \label{48)}
	A^{U} = \dfrac{B_{V}}{B_{R}}+5\dfrac{G_{V}}{G_{R}}-6
\end{equation}
According to the Ranganathan and Ostoja-Starzewski equation, Li et al.\cite{R86} suggested the following anisotropy index (A$^{R}$) for 2D materials:
\begin{equation} \label{49)}
	A^{R} = \dfrac{B_{V}}{B_{R}} + 2\dfrac{G_{V}}{G_{R}} - 3
\end{equation}
which B$_{V}$/B$_{R}$ and G$_{V}$/G$_{R}$ are area and shear modules in Voigt and Reuss approximations, respectively, which can be defined as follows \cite{R86}:
\begin{equation} \label{50)}
	B_{R} = \dfrac{1}{S_{11}+S_{22}+2S_{12}},
	B_{V} = \dfrac{C_{11}+C_{22}+2C_{12}}{4},
\end{equation}
\begin{equation} \label{51)}
	G_{R} = \dfrac{2}{S_{11}+S_{22}-2S_{12}+S_{66}},
	G_{V} = \dfrac{C_{11}+C_{22}-2C_{12}+4C_{66}}{8}.
\end{equation}
Zener proposed an anisotropy factor ($A^{Z}$) for crystals of cubic symmetry defined as the ratio of the extreme values of the orientation-dependent shear moduli given by \cite{R76}
\begin{equation} \label{52)}
	A^{Z} = \dfrac{2C_{44}}{C_{11}-C_{12}} 
\end{equation}

On the other hand, Chung and Buessem \cite{R70} observed that a (Cubic) crystal is isotropic when the Voigt average of the shear moduli $G_{V}$ over all possible orientations was equal to the inverse of the orientation averaged shear compliance (Reuss average)$G_{R}$, which motivated the adoption of the factor
\begin{equation} \label{53)} 
	A^{BC} = \dfrac{G_{V}-G_{R}}{G_{V}+G_{R}} 
\end{equation}
$A^{BC}$= 0 for isotropic materials, and any positive deviation from this limiting value would indicate an anisotropic behavior. With this deﬁnition, one can determine whether a given cubic crystal is more anisotropic than the other. This index is defined as follows:
\begin{equation} \label{54)} 
	A^{L} = \sqrt{ln(\dfrac{B_{V}}{B_{R}})^{2} +5 \,ln(\dfrac{G_{V}}{G_{R}})^{2}}
\end{equation}

The Kube's log-Euclidean anisotropy ($A^{L}$) is the most general deﬁnition of elastic anisotropy at present, as it was defined to make deﬁnitive comparisons between any two crystals. The isotropy is determined by $A^{L}$ = 0 and any positive value denotes a measure of the elastic anisotropy. 

Two similar anisotropy indices A$^{K}$ and A$^{SU}$ with the above equation, have been proposed by Li et al. for 2D materials as follows \cite{R86}:
\begin{equation} \label{55)} 
	A^{SU} = \sqrt{(\dfrac{{B_{V}}}{B_{R}}-1)^2+2(\dfrac{{G_{V}}}{G_{R}}-1)^2},  
\end{equation}
\begin{equation} \label{56)} 
	A^{K} = \sqrt{(ln \dfrac{{B_{V}}}{B_{R}})^2+2(ln \dfrac{{G_{V}}}{G_{R}})^2},  
\end{equation}
Since hardness is an essential property that is essential to describe the mechanical behavior fully various semi-empirical relations have been proposed to estimate hardness using the elastic moduli. In \textsc{ElaTools} package, the following semi-empirical correlations \cite{R78} between Vickers hardness ($H_{V}$) and B, G, E, $\nu$, and B/G, so-called macroscopic models for hardness prediction \cite{R79,R80,R81,R82}, are used:
\begin{equation} \label{57)} 
	H_{1a} = 0.0963B,  
\end{equation}
\begin{equation} \label{58)} 
	H_{1b} = 0.0607E,
\end{equation}
\begin{equation} \label{59)} 
	H_{2} = -2.899+0.1769G,
\end{equation}
\begin{equation} \label{60)} 
	H_{3} = 0.0635E,
\end{equation}
\begin{equation} \label{61)} 
	H_{4} =  \dfrac{B(1-2\nu)}{6(1+\nu)},
\end{equation}
\begin{equation} \label{62)} 
	H_{5} = 2(\dfrac{G^{2}}{B})^{0.585}-3.
\end{equation}
To determine the aptitude of the above methods in predicting hardness for different types of materials, we have used the model proposed by Singh et al. \cite{R55}. They found that the best model for these five hardness analysis methods correlates with the crystal class and the energy bandgap ($E_{g}$). Table \ref{Tab:1} provides a selection guide for the best method for calculating hardness for different types of compounds.

The longitudinal ($\nu_{l}$), transverse ($\nu_{t}$), and average ($\nu_{m}$) elastic wave velocities can be calculated from the knowledge of the B and
G, and $\rho$ as follows \cite{R83}:
\begin{equation} \label{63)} 
	\nu_{l} = \dfrac{3B+4G}{3\rho},
\end{equation}
\begin{equation} \label{64)} 
	\nu_{t} = \sqrt{\dfrac{G}{\rho}},
\end{equation}
\begin{equation} \label{65)} 
	\nu_{m} = [\dfrac{1}{3}(\dfrac{2}{\nu_{t}^3}+\dfrac{1}{\nu_{l}^3})]^{-1/3}.
\end{equation}
where G and B denote $G_{VRH}$ and $B_{VRH}$, respectively. Moreover, these equations imply that one can obtain the elastic moduli and elastic constants by measuring the elastic wave velocities using ultrasonic waves. 
\begin{table}[H]  
\caption{A guide to select the best hardness calculation method as a function of the crystal class and bandgap ($E_{g}$). This model was proposed by Singh et al. \cite{R55}} 
\centering
\label{Tab:1}
\begin{tabular}{cccccc}
	\hhline{======}
	\textbf{Type of material}                                                                                          & \textbf{Cubic} & \textbf{Hexagonal} & \textbf{Orthorhombic} & \textbf{Rhombohedral} & \textbf{General}  \\ 
	\hhline{======}
	\begin{tabular}[c]{@{}c@{}}\textbf{Insulator}\\\textbf{($E_{g}$ \textgreater{} 2 eV)}\end{tabular}               & $H_{2}$        & $H_{1b}$           & $H_{2}$               & $H_{2}$               & $H_{2}$           \\ 
	\hhline{------}
	\begin{tabular}[c]{@{}c@{}}\textbf{Semiconductor }\\\textbf{(0\textless{} $E_{g}$ \textless{} 2 eV)}\end{tabular} & $H_{5}$        & $H_{1b}$, $H_{3}$  & -                     & $H_{2}$               & $H_{5}$           \\ 
	\hhline{------}
	\begin{tabular}[c]{@{}c@{}}\textbf{Metal}\\\textbf{($E_{g}$ = 0)}\end{tabular}                                   & $H_{1a}$       & $H_{4}$           & $H_{4}$               & $H_{4}$               & $H_{4}$           \\
	\hhline{======}
\end{tabular}
\end{table}

\begin{table}[H] 
\caption{List of the main elastic (wave) properties and anisotropy indices of 2D and 3D materials.}
	\centering
	\label{Tab:2}
	\begin{tabular}{ccl} 
		\toprule
		 \hhline{===}
		\multicolumn{2}{c}{\textbf{Properties}}                                                                 & \textbf{\textbf{Formulae(s)}}  \\ 
		\hhline{===}
		\multirow{10}{*}{Elastic moduli and elastic parameters} & Bulk modulus (B)                               &      Eqs.\eqref{35)} and \eqref{36)}                       \\
		& Young's modulus (E)                             &       Eq.\eqref{37)}                      \\
		& Shear modulus (G)                               &       Eqs.\eqref{35)} and \eqref{36)}                      \\
		& P-wave modulus (M)                             &       Eq.\eqref{43)}                          \\
		& Poisson's ratio ($\nu$)                            &       Eqs.\eqref{37)} and \eqref{42)}                      \\
		& Pugh's ratio (B/G)                                 &    B/G                           \\
		& Lame's first parameter ($\lambda_{1}$)                      & Eq.\eqref{44)}                            \\
		&  Lame's second parameter ($\lambda_{2}$)&   Eq.\eqref{44)}                          \\
		& Kleinman's parameter ($\xi$)                       &   Eq.\eqref{45)}                          \\
		& Minimum thermal conductivity ($\kappa_{m}$)               &   Eq.\eqref{46)}                          \\ 
		\hhline{---}
		\multirow{3}{*}{Cauchy's pressures (P$_{C}$)}                       & Cubic symmetry                             &         Eq.\eqref{39)}                    \\
		& Hex., Trig., and Tetra. symmetries             &           Eq.\eqref{40)}                  \\
		& Orthorhombic symmetry                       &          Eq.\eqref{41)}                   \\ 
		\hhline{---}
		\multirow{7}{*}{Elastic anisotropy indices}               & Universal anisotropy index (A$^{U}$)                  &      Eq.\eqref{48)}                        \\
        & Zener's anisotropy index (A$^{Z}$)              &           Eq.\eqref{52)}                  \\
    	& Ranganathan anisotropy index (A$^{R}$)              &           Eq.\eqref{49)}                  \\
		& Chung-Buessem anisotropy index (A$^{CB}$)             &      Eq.\eqref{53)}                       \\
		& Kube's log-Euclidean anisotropy index (A$^{L}$)      &         Eq.\eqref{54)}                     \\ 
		& 2D anisotropy index (A$^{SU}$)              &           Eq.\eqref{55)}                  \\
	    & Kube anisotropy index (A$^{K}$)              &           Eq.\eqref{56)}                  \\
		\hhline{---}
		\multirow{6}{*}{Hardness methods}                        & H$_{1a}$                                    &  Eq.\eqref{57)}                           \\
		& H$_{1b}$                                    &  Eq.\eqref{58)}                           \\
		& H$_{2}$                                     & Eq.\eqref{59)}                            \\
		& H$_{3}$                                     & Eq.\eqref{60)}                            \\
		& H$_{4}$                                     & Eq.\eqref{61)}                            \\
		& H$_{5}$                                     & Eq.\eqref{62)}                            \\ 
		\hhline{---}
		\multirow{3}{*}{Elastic wave properties}                 & Longitudinal elastic wave velocity ($v_{l}$)         &   Eq.\eqref{63)}                          \\
		& Transverse elastic wave velocity ($v_{t}$)            & Eq.\eqref{64)}                             \\
		& Main~elastic wave velocity ($v_{m}$)                  &   Eq.\eqref{65)}                          \\   
		\bottomrule
		\hhline{===}
	\end{tabular}
\end{table}

\subsection{Appendix B} 
List of input, output, and temporary files in Table \ref{Tab:2}. The three executables \textsf{dat2gnu.x}, \textsf{dat2html.x}, and \textsf{dat2wrl.x} are called with input command “\textsf{pro}” (for 3D representations and 2D projections) and “\textsf{hmpro}” (for 2D head maps). The executable \textsf{dat2agr.x} runs two input commands, \textsf{box} (\textsf{boxpro}), and \textsf{polar} (\textsf{polarpro}) used for 2D projections in cartesian and polar coordinates, respectively. The full details of the input commands and the displayable features of each of these post-processing codes are listed in Tables \ref{Tab:32}, \ref{Tab:33}, and \ref{Tab:34}.

\begin{table}[H] 
	\centering
	\small
	\caption{List of input, output, and temporary files related to \textsf{Elatools.x}, \textsf{dat2gnu.x}, \textsf{dat2agr.x}, \textsf{dat2wrl.x}, and \textsf{dat2html.x} executables.}
	\label{Tab:31}
	\begin{threeparttable}
		\begin{tabular}{ccccc} 
			\hhline{=====}
			\textbf{Program} & \textbf{Input comment} & \textbf{Input file(s)}                                                                                                  & \textbf{Output file(s}                                                                                                                                            & \textbf{Temporary file(s)}                                                        \\ 
			\hhline{=====}
			\textbf{\textsc{El\textit{A}Tools}}    & -                      & \begin{tabular}[c]{@{}c@{}}Cij.dat,\\Cij-2D.dat,\\INVELC-matrix,\\elast.output,\\ELADAT,\\ElaStic\_2nd.out\end{tabular} & \begin{tabular}[c]{@{}c@{}}Sij.dat, DATA.dat,\\ 2dcut\_pro.dat \tnote{1} ,\\3d\_pro.dat,\\pro\_2d\_sys.dat\end{tabular} & \begin{tabular}[c]{@{}c@{}}HKL, MESH,\\.aelastpro,\\.MaMiout, \\ 3d\_SD.dat\end{tabular}  \\ 
			\hhline{-----}
			
			\textbf{dat2gnu} & pro, hmpro, phmpro \tnote{2}                 & \begin{tabular}[c]{@{}c@{}}2dcut\_pro.dat,\\HKL,\\.MaMiout,\\3d\_SD.dat\end{tabular}                                                            & gpi files                                                                                                                                                         & .SDdat                                                                                 \\ 
			\hhline{-----}
			\textbf{dat2agr} & polar,  box, boxpro, polarpro     & \begin{tabular}[c]{@{}c@{}}2dcut\_pro.dat\end{tabular}                                                           & agr files                                                                                                                                                         & -                                                                                 \\ 
			\hhline{-----}
			\textbf{dat2wrl} & pro                    & \begin{tabular}[c]{@{}c@{}}3d\_pro.dat,\\.aelastpro,\\.MaMiout\end{tabular}                                                        & wrl files                                                                                                                                                         & -                                                                                 \\
			\hhline{-----}
\textbf{dat2html} & pro                    & \begin{tabular}[c]{@{}c@{}}3d\_pro.dat\\MESH\end{tabular}                                                        & html files                                                                                                                                                         & -                                                                                \\			
			\hhline{=====}
		\end{tabular}
		\begin{tablenotes}
			\item[1]pro: bulk, comp, poisson, young, shear, pp, pf, ps, gp, gf, gs, pfp, pff, pfs, km, etc.
			\item[2] The full list of input comments is in Table \ref{Tab:32}, \ref{Tab:33}, and \ref{Tab:34}. Note that the current features and options of the \textsc{ElaTools} package may increase in future versions.
		\end{tablenotes}
	\end{threeparttable}
\end{table}

\begin{table}[H] 
	\centering
\caption{List of input commands, and input files related to \textsf{dat2gnu.x} executable.}
	\label{Tab:32}
	\begin{tabular}{cccc} 
		\hhline{====}
		\textbf{Input comment} & \multicolumn{1}{c}{\textbf{Input file}}                                         & \textbf{Property}    & \textbf{Type of graph}                                                                           \\ 
		\hhline{====}
		poi                    & 2dcut\_poisson.dat                                                                 & $\nu$                & \multirow{18}{*}{\begin{tabular}[c]{@{}c@{}}Polar coordinates\\ for 3D system\end{tabular}}      \\
		young                  & 2dcut\_young.dat                                                                   & E                    &                                                                                                  \\
		bulk                   & 2dcut\_bulk.dat                                                                    & B                    &                                                                                                  \\
		shear                  & 2dcut\_shear.dat                                                                   & G                    &                                                                                                  \\
		comp                   & 2dcut\_comp.dat                                                                    & $\beta$              &                                                                                                  \\
		pp                     & 2dcut\_pveloc.dat                                                                      & $\nu_{p}$: P-mode    &                                                                                                  \\
		ps                     & 2dcut\_pveloc.dat                                                                      & $\nu_{p}$: Show-mode &                                                                                                  \\
		pf                     & 2dcut\_pveloc.dat                                                                      & $\nu_{p}$: Fast-mode &                                                                                                  \\
		gp                     & 2dcut\_gveloc.dat                                                                      & $\nu_{g}$: P-mode    &                                                                                                  \\
		gs                     & 2dcut\_gveloc.dat                                                                      & $\nu_{g}$: Show-mode &                                                                                                  \\
		gf                     & 2dcut\_gveloc.dat                                                                      & $\nu_{g}$: Fast-mode &                                                                                                  \\
		pfp                    & 2dcut\_pfaveloc.dat                                                                     & PFA: P-mode          &                                                                                                  \\
		pfs                    & 2dcut\_pfaveloc.dat                                                                     & PFA: Show-mode       &                                                                                                  \\
		pff                    & 2dcut\_pfaveloc.dat                                                                     & PFA: Fast-mode       &                                                                                                  \\
		pall                   &  2dcut\_pveloc.dat     & $\nu_{p}$: All modes &                                                                                                  \\
		gall                   &  2dcut\_gveloc.dat    & $\nu_{g}$: All modes &                                                                                                  \\
		pfall                  & 2dcut\_pfaveloc.dat  & PFA: All modes       &                                                                                                  \\
		km                     & 3d\_km.dat                                                                      & $\kappa_{m}$         &                                                                                                  \\ 
		 \hhline{----}
		hmpoi                  & 3d\_poisson.dat                                                                 & $\nu$                & \multirow{15}{*}{\begin{tabular}[c]{@{}c@{}}Heat map diagram\\ for 3D system\end{tabular}}        \\
		hmyoung                & 3d\_young.dat                                                                   & E                    &                                                                                                  \\
		hmbulk                 & 3d\_bulk.dat                                                                    & B                    &                                                                                                  \\
		hmcomp                 & 3d\_comp.dat                                                                    & $\beta$              &                                                                                                  \\
		hmshear                & 3d\_bulk.dat                                                                    & G                    &                                                                                                  \\
		hmpall                 & \begin{tabular}[c]{@{}c@{}}3d\_pp.dat,\\3d\_ps.dat,\\3d\_pf.dat\end{tabular}    & $\nu_{p}$: All modes &                                                                                                  \\
		hmgall                 & \begin{tabular}[c]{@{}c@{}}3d\_gp.dat,\\3d\_gs.dat,\\3d\_gf.dat\end{tabular}    & $\nu_{g}$: All modes &                                                                                                  \\
		hmpfall                & \begin{tabular}[c]{@{}c@{}}3d\_pfp.dat,\\3d\_pfs.dat,\\3d\_pff.dat\end{tabular} & PFA: Fast-mode       &                                                                                                  \\
		hmkm                   & 3d\_km.dat                                                                      & $\kappa_{m}$         &                                                                                                  \\ 
	    \hhline{----}
		2dpoi                  & poisson\_2d\_sys.dat                                                            & $\nu$                & \multirow{3}{*}{\begin{tabular}[c]{@{}c@{}}Polar coordinates\\ for 2D system\end{tabular}}       \\
		2dyoung                & young\_2d\_sys.dat                                                              & E                    &                                                                                                  \\
		2dshear                & shear\_2d\_sys.dat                                                              & G                    &                                                                                                  \\ 
		\hhline{----}
		phmpoi                 & poisson\_2d\_sys.dat                                                            & $\nu$                & \multirow{3}{*}{\begin{tabular}[c]{@{}c@{}}Polar heat map diagram\\ for 2D system\end{tabular}}  \\
		phmyou                 & young\_2d\_sys.dat                                                              & E                    &                                                                                                  \\
		phmshe                 & shear\_2d\_sys.dat                                                              & G                    &                                                                                                  \\
		\hhline{====}
	\end{tabular}
\end{table}

\begin{table}[H] 
	\centering
\caption{List of input commands, and input files related to \textsf{dat2wrl.x} and \textsf{dat2html.x} executables.}
	\label{Tab:33}
	\begin{tabular}{cccc} 
		\hhline{====}
		\textbf{Input comment} & \textbf{Input file(s)}                                                       & \textbf{Property}    & \textbf{Type of graph}                                                                            \\ 
		\hhline{====}
		poi                    & 3d\_poisson.dat                                                              & $\nu$                & \multirow{21}{*}{\begin{tabular}[c]{@{}c@{}}Spherical coordinates \\ for 3D system\end{tabular}}  \\
		young                  & 3d\_young.dat                                                                & E                    &                                                                                                   \\
		bulk                   & 3d\_bulk.dat                                                                 & B                    &                                                                                                   \\
		shear                  & 3d\_shear.dat                                                                & G                    &                                                                                                   \\
		comp                   & 3d\_comp.dat                                                                 & $\beta$              &                                                                                                   \\
		pp                     & 3d\_pp.dat                                                                   & $\nu_{p}$: P-mode    &                                                                                                   \\
		ps                     & 3d\_ps.dat                                                                   & $\nu_{p}$: Show-mode &                                                                                                   \\
		pf                     & 3d\_pf.dat                                                                   & $\nu_{p}$: Fast-mode &                                                                                                   \\
		gp                     & 3d\_gp.dat                                                                   & $\nu_{g}$: P-mode    &                                                                                                   \\
		gs                     & 3d\_gs.dat                                                                   & $\nu_{g}$: Show-mode &                                                                                                   \\
		gf                     & 3d\_gf.dat                                                                   & $\nu_{g}$: Fast-mode &                                                                                                   \\
		pfp                    & 3d\_pfp.dat                                                                  & PFA: P-mode          &                                                                                                   \\
		pfs                    & 3d\_pfs.dat                                                                  & PFA: Show-mode       &                                                                                                   \\
		pff                    & 3d\_pff.dat                                                                  & PFA: Fast-mode       &                                                                                                   \\
		pall                   & \begin{tabular}[c]{@{}c@{}}3d\_pp.dat,\\3d\_ps.dat,\\3d\_pf.dat\end{tabular} & $\nu_{p}$: All modes &                                                                                                   \\
		gall                   & \begin{tabular}[c]{@{}c@{}}3d\_gp.dat,\\3d\_gs.dat,\\3d\_gf.dat\end{tabular} & $\nu_{g}$: All modes &                                                                                                   \\
		km                     & 3d\_km.dat                                                                   & $\kappa_{m}$         &                                                                                                   \\
		\hhline{====}
	\end{tabular}
\end{table}

\begin{table}[H] 
	\centering
\caption{List of input commands, and input files related to \textsf{dat2agr.x} executable.}
	\label{Tab:34}
	\begin{tabular}{cccc} 
		\hhline{====}
		\textbf{Input comment} & Input file(s)                                                                                                                       & \textbf{Property}                      & \textbf{Type of graph}                                                                                           \\ 
		\hhline{====}
		box                    & \begin{tabular}[c]{@{}c@{}}2dcut\_young.dat,\\2dcut\_shear.dat,\\2dcut\_buk.dat,\\2dcut\_comp.dat,\\2dcut\_poisson.dat\end{tabular} & E, G, B, $\beta$, and $\nu$: multiplot & \multirow{10}{*}{\begin{tabular}[c]{@{}c@{}}Polar coordinates \\of 2D cuts \\ in the 3D system\end{tabular}}     \\
		boxpoi                 & 2dcut\_poisson.dat                                                                                                                  & $\nu$                                  &                                                                                                                  \\
		boxyoung               & 2dcut\_young.dat                                                                                                                    & E                                      &                                                                                                                  \\
		boxbulk                & 2dcut\_buk.dat                                                                                                                      & B                                      &                                                                                                                  \\
		boxshear               & 2dcut\_shear.dat                                                                                                                    & G                                      &                                                                                                                  \\
		boxcomp                & 2dcut\_comp.dat                                                                                                                     & $\beta$                                &                                                                                                                  \\
		boxkm                  & 2dcut\_km.dat                                                                                                                       & $\kappa_{m}$                           &                                                                                                                  \\
		boxpall                & 2dcut pveloc.dat                                                                                                                    & $\nu_{p}$: All modes                   &                                                                                                                  \\
		boxgall                & 2dcut gveloc.dat                                                                                                                    & $\nu_{g}$: All modes                   &                                                                                                                  \\
		boxpfall               & 2dcut pdveloc.dat                                                                                                                   & PFA: All modes                         &                                                                                                                  \\ 
		\hhline{----}
		polar                  & \begin{tabular}[c]{@{}c@{}}2dcut\_young.dat,\\2dcut\_shear.dat,\\2dcut\_buk.dat,\\2dcut\_comp.dat,\\2dcut\_poisson.dat\end{tabular} & E, G, B, $\beta$, and $\nu$: multiplot & \multirow{10}{*}{\begin{tabular}[c]{@{}c@{}}Cartesian coordinates\\of 2D cuts \\ in the 3D system\end{tabular}}  \\
		polarpoi               & 2dcut\_poisson.dat                                                                                                                  & $\nu$                                  &                                                                                                                  \\
		polaryoung             & 2dcut\_young.dat                                                                                                                    & E                                      &                                                                                                                  \\
		polarbulk              & 2dcut\_buk.dat                                                                                                                      & B                                      &                                                                                                                  \\
		polarshear             & 2dcut\_shear.dat                                                                                                                    & G                                      &                                                                                                                  \\
		polarcomp              & 2dcut\_comp.dat                                                                                                                     & $\beta$                                &                                                                                                                  \\
		polarkm                & 2dcut\_km.dat                                                                                                                       & $\kappa_{m}$                           &                                                                                                                  \\
		polarpall              & 2dcut\_pveloc.dat                                                                                                                    & $\nu_{p}$: All modes                   &                                                                                                                  \\
		polargall              & 2dcut\_gveloc.dat                                                                                                                    & $\nu_{g}$: All modes                   &                                                                                                                  \\
		polarpfall             & 2dcut\_pfveloc.dat                                                                                                                   & PFA: All modes                         &                                                                                                                  \\
		\hhline{====}
	\end{tabular}
\end{table}
\subsection{Appendix C} 
List of main elastic properties and anisotropy indices of ZnAu$_{ 2}$(CN)$_{ 4}$, GaAs, CrB$_{2}$, $\delta$-phosphorene ($\delta$-P), and Pd$_{2}$O$_{6}$Se$_{2}$ compounds. The \textsc{El\textit{A}Tools} also calculates a measure of the anisotropy \textit{A}$_{M}$ of each elastic modulus \textit{M,} defined as follows: 
\begin{equation} \label{66)} 
A_{M}=\begin{cases}\frac{M_{MAX}}{M_{MIN}} & ;if\ sign(M_{MAX})=sign(M_{MAX}) \\ \infty  &\ ;otherwise\end{cases} 
\end{equation} 

$A_{M}$ is particularly interesting as the marked anisotropy of the mechanical properties is often associated with anomalous mechanical behavior, such as NPR and NLC. As can be seen in Table IV, when $A_{M}$ is infinite, the material has anomalous mechanical properties.
\begin{table}[H]   
	\centering
		\label{Tab:3}
	\caption{The main elastic properties of ZnAu$_{2}$(CN)$_{4}$ compound.}
	\begin{tabular}{cccccccc} 
		\hhline{========}
		\begin{tabular}[c]{@{}c@{}}\textbf{Elastic}\\\textbf{properties}\end{tabular} & \begin{tabular}[c]{@{}c@{}}\textbf{Bulk}\\\textbf {modulus}\\\textbf{(GPa)}\end{tabular} & \begin{tabular}[c]{@{}c@{}}\textbf{Shear}\\\textbf{modulus}\\\textbf{(GPa)}\end{tabular} & \begin{tabular}[c]{@{}c@{}}\textbf{Young}\\\textbf{modulus}\\\textbf{(GPa)}\end{tabular} & \begin{tabular}[c]{@{}c@{}}\textbf{Poisson’s}\\\textbf{ratio}\end{tabular} & \begin{tabular}[c]{@{}c@{}}\textbf{Pugh}\\\textbf{ratio}\end{tabular} & \begin{tabular}[c]{@{}c@{}}\textbf{P-wave}\\\textbf{modulus}\\\textbf{(GPa)}\end{tabular} & \begin{tabular}[c]{@{}c@{}}\textbf{Linear}\\\textbf{Compressibility}\\\textbf{(TPa$^{-1}$)}\end{tabular}  \\ 
		\hhline{========}
		\textbf{Max}                                                                  & 5.747×10$^{2}$                                                  & 12.10                                                                                    & 28.25                                                                                    & 1.255                                                                      & -                                                                     & -                                                                                         & 62.328                                                                                                             \\ 
		\hhline{--------}
		\textbf{Min}                                                                  & -5.477×10$^{2}$                                                & 3.18                                                                                     & 7.26                                                                                     & -0.021                                                                     & -                                                                     & -                                                                                         & -51.689                                                                                                            \\ 
		\hhline{--------}
		\textbf{Voigt}                                                                & 55.756                                                                        & 8.753                                                                                    & 24.954                                                                                   & 0.4254                                                                     & 6.3696                                                                & 67.4267                                                                                   & -                                                                                                                  \\ 
		\hhline{--------}
		\textbf{Reuss}                                                                & 13.705                                                                        & 4.618                                                                                    & 12.456                                                                                   & 0.4597                                                                     & 2.9674                                                                & 19.8627                                                                                   & -                                                                                                                  \\ 
		\hhline{--------}
		\begin{tabular}[c]{@{}c@{}}\textbf{Average}\\\textbf{(Hall)}\end{tabular}     & 34.730                                                                        & 6.686                                                                                    & 18.705                                                                                   & 0.4426                                                                     & 5.1945                                                                & 43.6447                                                                                   & -                                                                                                                  \\ 
		\hhline{--------}
		\textbf{$\textbf{A}_{\textbf{M}}$}                                                           & $\infty$~                                                        & 3.804                                                                                    & 3.893                                                                                    & $\infty$~                                                     & -                                                                     & -                                                                                         & $\infty$~                                                                                             \\
		\hhline{========}
	\end{tabular}
\end{table}

\begin{table}[H] 
	\centering
		\label{Tab:4}
	\caption{The main elastic properties of the CrB$_{2}$ compound.\\}
	\begin{tabular}{cccccccc} 
		\hhline{========}
		\begin{tabular}[c]{@{}c@{}}\textbf{Elastic}\\\textbf{properties}\end{tabular} & \begin{tabular}[c]{@{}c@{}}\textbf{Bulk }\\\textbf{modulus}\\\textbf{(GPa)}\end{tabular} & \begin{tabular}[c]{@{}c@{}}\textbf{Shear}\\\textbf{modulus}\\\textbf{(GPa)}\end{tabular} & \begin{tabular}[c]{@{}c@{}}\textbf{Young}\\\textbf{modulus}\\\textbf{(GPa)}\end{tabular} & \begin{tabular}[c]{@{}c@{}}\textbf{Poisson’s}\\\textbf{ratio}\end{tabular} & \begin{tabular}[c]{@{}c@{}}\textbf{Pugh}\\\textbf{ratio}\end{tabular} & \begin{tabular}[c]{@{}c@{}}\textbf{P-wave}\\\textbf{modulus}\\\textbf{(GPa)}\end{tabular} & \begin{tabular}[c]{@{}c@{}}\textbf{Linear}\\\textbf{Compressibility}\\\textbf{(TPa$^{-1}$)}\end{tabular}  \\ 
		\hhline{========}
		\textbf{Max}                                                                  & 11.922×10$^{2}$                                                            & 197.40                                                                                   & 463.18                                                                                   & 0.453                                                                      & -                                                                     & -                                                                                         & 1.869                                                                                                              \\ 
		\hhline{--------}
		\textbf{Min}                                                                  & 5.351×10$^{2}$                                                              & 125.04                                                                                   & 260.22                                                                                   & 0.162                                                                      & -                                                                     & -                                                                                         & 0.839                                                                                                              \\ 
		\hhline{--------}
		\textbf{Voigt}                                                                & 295.122                                                                                  & 169.833                                                                                  & 427.497                                                                                  & 0.2586                                                                     & 1.7377                                                                & 521.5667                                                                                  & -                                                                                                                  \\ 
		\hhline{--------}
		\textbf{Reuss}                                                                & 282.009                                                                                  & 161.560                                                                                  & 406.966                                                                                  & 0.2595                                                                     & 1.7455                                                                & 497.4226                                                                                  & -                                                                                                                  \\ 
		\hhline{--------}
		\begin{tabular}[c]{@{}c@{}}\textbf{Average}\\\textbf{(Hall)}\end{tabular}     & 288.565                                                                                  & 165.697                                                                                  & 417.231                                                                                  & 0.2635                                                                     & 1.7415                                                                & 509.4947                                                                                  & -                                                                                                                  \\ 
		\hhline{--------}
		\textbf{$\textbf{A}_{\textbf{M}}$}                                                           & 2.227                                                                                   & 1.579                                                                                    & 1.780                                                                                    & 2.793                                                                      & -                                                                     & -                                                                                         & 2.228                                                                                                              \\
		\hhline{========}
	\end{tabular}
\end{table}

\begin{table}[H]  
	\centering

	\caption{List of Maximum (Max), minimum (Min) phase and group velocities, and anisotropy values of GaAs compound.}
		\label{Tab:8}
	\begin{tabular}{ccccccc} 
		\hhline{=======}
		\multirow{2}{*}{\textbf{Property }} & \multicolumn{3}{c}{\textbf{Phase velocity (km/s)}}                    & \multicolumn{3}{c}{\textbf{Group velocity (km/s)}}                      \\ 
		\cline{2-7}
		& \textbf{P mode} & \textbf{FS mode} & \textbf{SS mode} & \textbf{P mode} & \textbf{FS mode} & \textbf{SS mode}  \\ 
		\hhline{=======}
		\textbf{Max}                        & 5.398            & 3.346                    & 3.346                    & 5.398            & 3.369                    & 3.490                     \\
		\textbf{Min}                        & 4.731            & 2.805                    & 2.475                    & 4.731            & 2.935                    & 2.476                     \\
		\textbf{A$_{M}$}                             & 1.14             & 1.19                     & 1.35                     & 1.14             & 1.15                     & 1.41                      \\
		\hhline{=======}
	\end{tabular}
\end{table}

\begin{table}[H]   
	\centering
		\label{Tab:5}
	\caption{The main elastic properties of the $\delta$-P 2D compound. \\}
	\begin{tabular}{ccccc} 
		\hhline{=====}
		\begin{tabular}[c]{@{}c@{}}\textbf{Elastic}\\\textbf{properties}\end{tabular} & \begin{tabular}[c]{@{}c@{}}\textbf{Area modulus}\\\textbf{(N/m)}\end{tabular} & \begin{tabular}[c]{@{}c@{}}\textbf{Shear modulus}\\\textbf{(N/m)}\end{tabular} & \begin{tabular}[c]{@{}c@{}}\textbf{Young modulus}\\\textbf{(N/m)}\end{tabular} & \begin{tabular}[c]{@{}c@{}}\textbf{Poisson’s}\\\textbf{ratio}\end{tabular}  \\ 
		\hhline{=====}
		\textbf{Max}                                                                  & -                                                                            & 66.452                                                                         & 142.877                                                                        & 0.290                                                                       \\ 
		\hhline{-----}
		\textbf{Min}                                                                  & -                                                                            & 24.500                                                                         & 62.277                                                                         & -0.267                                                                      \\ 
		\hhline{-----}
		\textbf{Voigt}                                                                & 47.608                                                                       & 47.909                                                                         & -                                                                              & -                                                                           \\ 
		\hhline{-----}
		\textbf{Reuss}                                                                & 44.395                                                                       & 35.808                                                                         & -                                                                              & -                                                                           \\ 
		\hhline{-----}
		\textbf{\textit{xy}-plane}                                                    & -                                                                            & 24.500                                                                         & -                                                                              & -0.159                                                                      \\ 
		\hhline{-----}
		\textbf{\textit{yx}-plane}                                                    & -                                                                            & -                                                                              & -                                                                              & -0.267                                                                      \\ 
		\hhline{-----}
		\textbf{\textit{x}-direction}                                                 & -                                                                            & -                                                                              & 84.872                                                                         & -                                                                           \\ 
		\hhline{-----}
		\textbf{\textit{y}-direction}                                                 & -                                                                            & -                                                                              & 142.868                                                                        & -                                                                           \\
		\hhline{=====}
	\end{tabular}
\end{table}

\begin{table}[H]   
\centering
\caption{The main elastic properties of the Pd$_{2}$O$_{6}$Se$_{2}$ 2D compound. \\}
	\label{Tab:9}
	\begin{tabular}{ccccc} 
		\hhline{=====}
		\begin{tabular}[c]{@{}c@{}}\textbf{Elastic}\\\textbf{properties}\end{tabular} & \begin{tabular}[c]{@{}c@{}}\textbf{Area modulus}\\\textbf{(N/m)}\end{tabular} & \begin{tabular}[c]{@{}c@{}}\textbf{Shear modulus}\\\textbf{(N/m)}\end{tabular} & \begin{tabular}[c]{@{}c@{}}\textbf{Young modulus}\\\textbf{(N/m)}\end{tabular} & \begin{tabular}[c]{@{}c@{}}\textbf{Poisson's}\\\textbf{ratio}\end{tabular}  \\ 
		\hhline{=====}
		\textbf{Max}                                                                  & -                                                                            & 16.682                                                                         & 65.892                                                                        & 1.315                                                                       \\ 
		\hhline{-----}
		\textbf{Min}                                                                  & -                                                                            & 4.857                                                                         & 6.573                                                                        & -0.492                                                                      \\ 
		\hhline{-----}
		\textbf{Voigt}                                                                & 23.445                                                                       & 15.828                                                                         & -                                                                              & -                                                                           \\ 
		\hhline{-----}
		\textbf{Reuss}                                                                & 7.686                                                                       & 7.524                                                                         & -                                                                              & -                                                                           \\ 
		\hhline{-----}
		\textbf{\textit{xy}-plane}                                                    & -                                                                            & 14.930                                                                         & -                                                                              & 0.168                                                                      \\ 
		\hhline{-----}
		\textbf{\textit{yx}-plane}                                                    & -                                                                            & -                                                                              & -                                                                              & 0.164                                                                      \\ 
		\hhline{-----}
		\textbf{\textit{x}-direction}                                                 & -                                                                            & -                                                                              & 39.731                                                                         & -                                                                           \\ 
		\hhline{-----}
		\textbf{\textit{y}-direction}                                                 & -                                                                            & -                                                                              & 38.390                                                                        & -                                                                           \\
		\hhline{=====}
	\end{tabular}
\end{table}

\begin{table}[H]  
	\centering
		\label{Tab:6}
	\caption{List of anisotropy indices of ZnAu$_{2}$(CN)$_{4}$, CrB$_{2}$, $\delta$-phosphorene ($\delta$-P), and Pd$_{2}$O$_{6}$Se$_{2}$ compounds.}
\begin{tabular}{ccccc} 
	\hhline{=====}
	\multirow{2}{*}{\textbf{Anisotropy index}} \\ \multirow{2}{*}{\textbf{and Cauchy pressure}} & \multicolumn{4}{c}{\textbf{Compounds }}                                                                                                                   \\ 
	\cline{2-5}
	& \textbf{\textbf{\textbf{\textbf{ZnAu$_{2}$(CN)$_{4}$}}}} & \textbf{\textbf{CrB$_{2}$}} & \textbf{\textbf{$\delta$-P}} & \textbf{Pd$_{2}$O$_{6}$Se$_{2}$}  \\ 
	\hhline{=====}
	\textbf{\textbf{A$^{U}$}}                           & 7.5448                                                   & 0.3025                      & -                            & -                                 \\ 
	\hhline{-----}
	\textbf{\textbf{A$^{L}$}}                           & 4.7370                                                   & 0.3025                      & -                            & -                                 \\ 
	\hhline{-----}
	\textbf{\textbf{A$^{CB}$}}                          & 0.3092                                                   & 0.0250                      & -                            & -                                 \\ 
	\hhline{-----}
	\textbf{\textbf{A$^{SU}$}}                          & -                                                        & -                           & 0.4833                      & 2.5760                            \\ 
	\hhline{-----}
	\textbf{\textbf{A$^{R}$}}                           & -                                                        & -                           & 0.7482                       & 4.2574                            \\ 
	\hhline{-----}
	\textbf{\textbf{A$^{K}$}}                           & -                                                        & -                           & 0.1814                       & 0.6657                            \\ 
	\hhline{-----}
	\textbf{\textbf{P$_{C}^{a}$}}                       & 48.50                                                    & 25.90                       &                              & -                                 \\ 
	\hhline{-----}
	\textbf{\textbf{P$_{C}^{c}$}}                       & 26.50                                                    & -15.60                      &                              & -                                 \\
	\hhline{=====}
\end{tabular}
\end{table}

\subsection{Appendix D}
Colors available in the visualization of elastic properties in the \textsc{El\textit{A}Tools}. Personalization of colors is provided in the Supplementary Information file.
\begin{table}[H]  
		\label{Tab:7}
	\centering
	\caption{List of default colors in the 3D and 2D visualization of elastic properties such as Young's modulus, bulk modulus, shear modulus, linear compressibility, and Poisson's ratio.}
	\begin{tabular}{cccc} 
		\hhline{====}
		\begin{tabular}[c]{@{}c@{}}\textbf{3D or 2D representation}\\\textbf{of elastic proprieties}\end{tabular} & \begin{tabular}[c]{@{}c@{}}\textbf{Positive value or}\\\textbf{Maximum positive}\\\textbf{value}\end{tabular} & \begin{tabular}[c]{@{}c@{}}\textbf{Minimum positive}\\\textbf{value}\end{tabular} & \begin{tabular}[c]{@{}c@{}}\textbf{Negative value or}\\\textbf{Minimum Negative}\\\textbf{value}\end{tabular}  \\ 
		\hhline{====}
		\textbf{Young’s modulus}                                                                                  & green                                                                                                         & -                                                                                 & -                                                                                                              \\ 
		\hhline{----}
		\textbf{Bulk modulus}                                                                                     & green                                                                                                         & -                                                                                 & -                                                                                                              \\ 
		\hhline{----}
		\textbf{Shear modulus}                                                                                    & bule                                                                                                          & green                                                                             & -                                                                                                              \\ 
		\hhline{----}
		\textbf{Linear compressibility}                                                                           & green                                                                                                         & -                                                                                 & red                                                                                                            \\ 
		\cline{2-4}
		\hhline{----}
		\textbf{Poisson’s ratio}                                                                                  & blue                                                                                                          & green                                                                             & red                                                                                                            \\
		\hhline{====}
	\end{tabular}
\end{table}

\section{acknowledgement}
Parviz Saeidi is acknowledged for the valuable comments on the first draft of the manuscript.
\section{References}

	\bibliographystyle{apsrev4-1} 
	\bibliography{ref.bib}


\end{document}